\documentclass[12pt]{article}

\usepackage[all]{xy}

\author{Yu.~M.~Zinoviev
       \thanks{E-mail address: Yurii.Zinoviev@ihep.ru} \\[0.5cm]
        {\it Institute for High Energy Physics} \\
        {\it of National Research Center "Kurchatov Institute"} \\
        {\it Protvino, Moscow Region, 142280, Russia}}

\title{On massive higher spin supermultiplets in $d=4$}

\date{}

\begin{document}

\maketitle

\begin{abstract}
In this work we discuss the cubic interactions for massless spin 3/2
gravitino with massive higher spin supermultiplets using three
superblocks $(2,3/2)$, $(5/2,2)$ and $(3,5/2)$ as the first
non-trivial examples. We use gauge invariant formalism for the massive
higher spin fields and, as it common in such cases, we face an
ambiguity related with the possible field redefinitions due to
the presence of Stueckelberg fields. From one hand, we show how this
ambiguity can be used as one more way to classify possible cubic
vertices. We also note that all these field redefinitions do not
change the part of the Lagrangian which appears in the unitary gauge
(where all Stueckelberg fields are set to zero) so we still have some
important independent results. From the other hand, we show how using
the so-called unfolded formalism one can fix these ambiguities and
obtain consistent deformations for all massive field gauge invariant
curvatures which is the most important step in the Fradkin-Vasiliev
formalism. Unfortunately, this works for the massive fields only so
the
way to construct deformations for the massless field curvatures is
still has to be found.
\end{abstract}

\thispagestyle{empty}
\newpage
\setcounter{page}{1}

\section{Introduction}

It is very well known that in $N=1$ $d=4$ supersymmetry there exist
two types of massless supermultiplets each containing one boson and
one
fermion which differ in spins by 1/2 (see e.g. 
\cite{Cur79,Vas80,KSP93,KS93,KS94} and references therein):
$$
\xymatrix{
H_{s+1} \ar@{<->}[r] & \Phi_{s+1/2} & & 
\Phi_{s+1/2} \ar@{<->}[r] &  H_s  }
$$

A classification of the cubic
interactions for arbitrary three massless $N=1$ supermultiplets in
light-cone formalism has been constructed in \cite{Met19a} (see also
\cite{Met19b} for extended supersymmetry). In the covariant formalism
there exists a number of explicit examples which are of two types:
\begin{itemize}
\item abelian \cite{BGK17,BGK18,BGK18a,BGK18b,GK19} 
$$
{\cal L}_1 \sim \Phi_1 D \Phi_2 D \Phi_3,
$$
where two lowest (super)spins enter through their gauge invariant
field strengths;
\item non-abelian \cite{KhZ20b}
$$
{\cal L}_1 \sim \Phi_1 \Phi_2 D \Phi_3,
$$
where only lowest (super)spin enter through its gauge invariant field
strength.
\end{itemize}
Recall that in general non-abelian vertices correspond  to the
situations when spins satisfy a so called strict triangular inequality
$s_1 < s_2 + s_3$ (assuming $s_1 \ge s_2 \ge s_3$).

Global supertransformations for the massless supermultiplets leaving a
sum of two free Lagrangians invariant are very well known (both in
components as well as in terms of superfields). To make the
supertransformations local one has to construct an interaction of
such supermultiplets with $N=1$ supergravity. In the first non-trivial
order one has to solve two separate tasks: cubic interaction with
massless spin 2 graviton and cubic interaction with massless spin 3/2
gravitino. Cubic interaction vertices for such supermultiplets with
massless spin 3/2 gravitino $(s_1,s_2,3/2)$, $s_1 - s_2 = 1/2$ (in the
same frame-like multispinor formalism we use here) can  be found in
\cite{KhZ20a}. They have $n = s_1 + s_2 - 5/2$ derivatives and are of
non-minimal type
$$
{\cal L}_1 \sim \Phi_1 \Phi_2 D \Psi,
$$
where gravitino enters through its gauge invariant field strength
only so that we do not have any non-trivial supertransformations.

As in the case of the gravitational interactions, there are two
possibilities to obtain minimal interaction with the standard
supertransformations. One possibility is to introduce a non-zero
cosmological constant $\Lambda < 0$ and work in the $AdS_4$ space. In
this case the vertex shown above acquires a whole tail of lower
derivative terms with the coefficients proportional to positive
degrees of the cosmological constant and among them one finds the
desired minimal interactions. Another possibility is to introduce 
non-zero masses for the supermultiplets and this is a subject of our
work here.

As is well known, there also exist two types of the massive
supermultiplets each containing four fields (two bosons and two
fermions):
$$
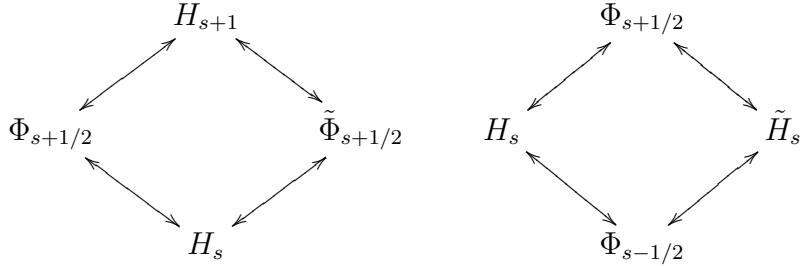
\xymatrix{
  & H_{s+1} \ar@{<->}[dl] &  &   & \Phi_{s+1/2} \ar@{<->}[dl] & \\
\Phi_{s+1/2} \ar@{<->}[dr] & & \tilde\Phi_{s+1/2} \ar@{<->}[ul] &
 H_s \ar@{<->}[dr] & & \tilde{H}_s \ar@{<->}[ul]  \\
  & H_s \ar@{<->}[ur] &    &  & \Phi_{s-1/2} \ar@{<->}[ur] }
$$
Contrary to the massless case, a construction of the massive
supermultiplet (i.e. to find global supertransformations leaving a sum
of four free Lagrangians invariant and such that superalgebra closes)
appears to be highly non-trivial. Till now the most effective way is
based on the gauge invariant formulations for the massive bosonic and
fermionic higher spin fields \cite{Zin01,Met06,Zin08b,PV10,KhZ19}
\footnote{For some results in the superfield formalism see
\cite{BGPL02,BGLP02,Kou20}}.
In the metric like formalism the massive arbitrary spin
supermultiplets were constructed in \cite{Zin07a} while their
realization in the frame-like formalism was elaborated in
\cite{BKhSZ19} (see also \cite{BKhSZ19a,BKhSZ19b} for the
supermultiplets containing partially massless and infinite spin
fields). Recall that in the gauge invariant formalism a massive field
is constructed out of the appropriate set of the massless ones. In
these the structure of the supertransformations schematically looks
like:
$$
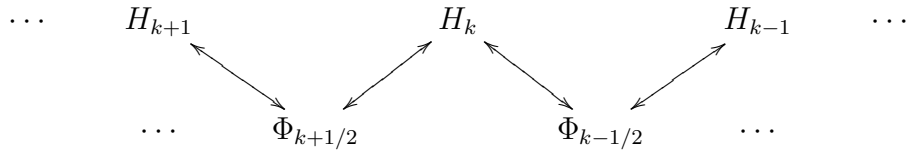
\xymatrix{
\cdots &  H_{k+1} \ar@{<->}[dr] &   &  H_k \ar@{<->}[dr] &  & H_{k-1}
& \cdots \\
 & \cdots &  \Phi_{k+1/2} \ar@{<->}[ur] &  &  \Phi_{k-1/2} 
\ar@{<->}[ur] & \cdots & }
$$
Thus each helicity component is connected by supersymmetry with two
closest ones while the explicit form of these supertransformations is
similar to the one in the massless case (up to some corrections
proportional to the mass). In this, the relative coefficients for
these supertransformations are fixed by the gauge invariance of the
massive fields.

Having the free massive supermultiplets constructed, the next
important task is to consider their interaction with $N=1$
supergravity. As in the massless case, in the first non-trivial
order (i.e. for the cubic interactions) this task can be divided into
two separate ones: interaction with massless spin 2 graviton and
massless spin 3/2 gravitino. Moreover, one can start just with the
pair $(s+1,s+1/2)$ or $(s+1/2,s)$ which we call superblocks in
\cite{BKhSZ19}. The aim of the current work is to consider
interactions for the massless spin 3/2 gravitino and three massive
superblocks $(2,3/2)$, $(5/2,2)$ and $(3,5/2)$ as the first
non-trivial examples.

Let us make some remarks on the general structure of such vertices.
Having the global supertransformations leaving the sum of the free
Lagrangians invariant, we can turn to the local ones $\eta \Rightarrow
\eta(x)$. In this, the sum of the free Lagrangians cease to be
invariant and variations have the form $\delta {\cal L}_0 \sim J
D\eta$ where $J$ --- some conserved supercurrent. As usual, we can
compensate this terms by adding appropriate interactions of the form 
${\cal L}_1 \sim J \Psi$. But if supercurrent is constructed out of
the higher spin fields which are gauge fields themselves, we must add
some non-minimal terms to make the vertex to be gauge invariant. Thus
we expect that in general the vertex will contain some combination of
minimal and non-minimal interactions
$$
{\cal L}_1 \sim J \Psi + \tilde{J} D \Psi.
$$

In the frame-like formalism an effective way to construct cubic
interaction vertices is the so-called Fradkin-Vasiliev formalism
\cite{FV87,FV87a,Vas11}. Let us briefly recall its main steps.
\begin{itemize}
\item The free Lagrangian for each field can be rewritten in the 
explicitly gauge invariant form
$$
{\cal L}_0 \sim \sum {\cal R} {\cal R},
$$,,
where ${\cal R} $ is a complete set of gauge invariant objects
(curvatures).
\item One consider the most general quadratic deformations for all
curvatures $\tilde{\cal R} = {\cal R} + \Delta {\cal R}$:
$$
\Delta {\cal R} \sim \Phi \Phi \quad \Leftrightarrow \quad
\delta_1 \Phi \sim \Phi \xi,
$$
which simultaneously determines all the corrections to the gauge
transformations.
\item The main consistency requirement is that all deformed curvatures
must transform covariantly
$$
\delta \hat{\cal R} \sim {\cal R} \xi.
$$
\item At last, the interacting Lagrangian takes the form
$$
{\cal L} \sim \sum \hat{\cal R} \hat{\cal R} +
\sum {\cal R} {\cal R} \Phi,
$$
i.e. the free Lagrangian where all curvatures are replaced by the
deformed ones plus possible abelian vertices.
\end{itemize}
For the massless fields (and massless supermultiplets) this formalism
nicely work without ambiguities. But in the massive case (due to the
presence of Stueckelberg fields) one face the problem related with
possible field redefinitions. First of all, remind that in general one
can distinguish three types of cubic vertices:
\begin{itemize}
\item trivially gauge invariant ones
$$
{\cal L}_1 \sim {\cal R} {\cal R} {\cal R} \quad \Rightarrow \quad
\delta_1 \phi \sim 0, \qquad [\delta_1, \delta_2] = 0,
$$
constructed out of the gauge invariant objects and do not modify
gauge transformations;
\item abelian ones
$$
{\cal L}_1 \sim {\cal R} {\cal R} \Phi \quad \Rightarrow \quad
\delta_1 \Phi \sim {\cal R} \xi, \qquad [\delta_1, \delta_2 ] = 0,
$$
which requires corrections to the gauge transformations but the
algebra remains to be abelian;
\item non-abelian ones
$$
{\cal L}_1 \sim {\cal R} \Phi \Phi \quad \Rightarrow \quad
\delta_1 \Phi \sim \Phi \xi, \qquad [\delta_1, \delta_2 ] \ne 0,
$$
which modify both gauge transformations and the gauge algebra.
\end{itemize}
The main results of the general analysis elaborated in \cite{BDGT18}
can be formulated as follows.
\begin{enumerate}
\item Any cubic vertex for three massive fields in the gauge invariant
formalism by the field (and gauge parameters) redefinitions can be
transformed into the abelian form. In the Fradkin-Vasiliev formalism
it means
$$
\Delta {\cal R} \sim {\cal B} \Phi \quad \Rightarrow \quad
\delta_1 \Phi \sim {\cal B} \xi, 
$$
where ${\cal B}$ are the gauge invariant one-forms for the
Stueckelberg fields.
\item By further (having more derivatives) field redefinitions this
vertex can be reduced to the trivially gauge invariant form, i.e.
$$
\Delta {\cal R} \sim {\cal B} {\cal B} \quad \Rightarrow \quad
\delta_1 \Phi = 0. 
$$
\end{enumerate}
Note that in \cite{BDGT18} only the case with three massive fields was
considered, while all our experience in application of the
Fradkin-Vasiliev formalism shows that when one has one massless and
two massive fields only the first statement holds (see for example
recent works \cite{KhZ21,KhZ21a}). Thus it seems that for any vertex
we have a whole family of formally equivalent representations
connected by the field redefinitions. Let us stress that due to the
higher derivative nature of these redefinitions it does not mean that
they are physically equivalent because a number of the physical
degrees
of freedom may be different. In our opinion, the best strategy is to
choose representatives with a minimum number of derivatives though
technically it is not easy to realize. 

For what follows it is important to note that any such redefinitions
always contain Stueckelberg fields and so do not change a part of the
vertex which appears in the so-called unitary gauge (when all
Stueckelberg fields are set to zero). Thus we still have some
physically important results independent of the field redefinitions. 

As it was already mentioned our aim here is to consider interactions
of the massless spin 3/2 gravitino with three massive superblocks 
$(2,3/2)$, $(5/2,2)$ and $(3,5/2)$. In Section 2 we provide all
necessary kinematical information on the gauge invariant description
of massive spin 3/2, 2, 5/2 and 3 fields. In Section 3 we consider the
most general abelian vertices and require them to be gauge invariant.
In all three cases we found that when the bosonic and fermionic
masses are different all the solutions (their number grows with spins)
correspond to some trivially gauge invariant vertices, while for
equal masses there exists just one non-trivial solution which can be
considered as a candidate for the minimal vertex. In Sections 4, 5 and
6 we show how using the so-called unfolded formalism 
\cite{PV10,PV10a,MV13,KhZ19,KhZ20} one
can fix the ambiguity related with the field redefinitions. The main
idea is to consider consistent deformations of the unfolded equations
for the massive boson and fermion in the presence of massless spin 3/2
gravitino. The results thus obtained can be used as the deformation of
curvatures for the massive fields in the Fradkin-Vasiliev formalism.
Unfortunately, this still leaves us with the ambiguities for the
massless spin 3/2 deformations which requires some additional ideas.

\section{Kinematics}

In this section we provide all necessary kinematical information on
the gauge invariant description for massive spin 3/2, 2, 5/2 and 3
fields. Complete description for arbitrary spin massive bosons and
fermions can be found in \cite{KhZ19}. We work in the frame-like
multispinor formalism where all objects are forms having some number
of (separately symmetric) dotted and undotted spinor indices
$\alpha,\dot\alpha = 1,2$.

\subsection{Spin 3/2}

In this case we need only physical fields: one-forms $\Phi^\alpha$, 
$\Phi^{\dot\alpha}$ and Stueckelberg zero-forms $\phi^\alpha$, 
$\phi^{\dot\alpha}$. Their gauge invariant curvatures look like:
\begin{eqnarray}
{\cal F}^\alpha &=& D \Phi^\alpha + \tilde{M} e^\alpha{}_{\dot\alpha}
\Phi^{\dot\alpha} - \frac{a_0}{3} E^\alpha{}_\beta \phi^\beta,
\nonumber \\
{\cal C}^\alpha &=& D \phi^\alpha - a_0 \Phi^\alpha + \tilde{M}
e^\alpha{}_{\dot\alpha} \phi^{\dot\alpha}, 
\end{eqnarray}
where
$$
a_0{}^2 = 6\tilde{M}^2,
$$
while basic two-forms $E^{\alpha(2)}$ and $E^{\dot\alpha(2)}$ are
defined as follows:
$$
e^{\alpha\dot\alpha} \wedge e^{\beta\dot\beta} =
\epsilon^{\dot\alpha\dot\beta} E^{\alpha\beta} +
\epsilon^{\alpha\beta} E^{\dot\alpha\dot\beta}. 
$$
These curvatures satisfy the following differential identities, which
play an important role in what follows:
\begin{eqnarray}
D {\cal F}^\alpha &=& - \tilde{M} e^\alpha{}_{\dot\alpha}
{\cal F}^{\dot\alpha} - \frac{a_0}{3} E^\alpha{}_\beta
{\cal C}^\beta, \nonumber \\
D {\cal C}^\alpha &=& - a_0 {\cal F}^{\alpha} - \tilde{M}
e^\alpha{}_{\dot\alpha} {\cal C}^{\dot\alpha}. \label{dif_1}
\end{eqnarray}
 We also need the so-called unfolded equations for the physical fields
\cite{KhZ20}:
\begin{eqnarray}
0 &=& D \Phi^\alpha + \tilde{M} e^\alpha{}_{\dot\alpha}
\Phi^{\dot\alpha} - \frac{a_0}{3} E^\alpha{}_\beta \phi^\beta +
E_{\beta(2)} \phi^{\alpha\beta(2)}, \nonumber \\
0 &=& D \phi^\alpha - a_0 \Phi^\alpha + \tilde{M} 
e^\alpha{}_{\dot\alpha} \phi^{\dot\alpha} + e_{\beta\dot\alpha}
\tilde\phi^{\alpha\beta\dot\alpha}. 
\end{eqnarray}
Here zero-forms $\phi^{\alpha(3)}$ and 
$\tilde\phi^{\alpha(2)\dot\alpha}$ are the first two representatives
of two infinite sets of gauge invariant zero-forms. Their equations 
$(k \ge 0)$:
\begin{eqnarray}
0 &=& D \phi^{\alpha(3+k)\dot\alpha(k)} + e_{\beta\dot\beta}
\phi^{\alpha(3+k)\beta\dot\alpha(k)\dot\beta} + c_{3,k}
e^\alpha{}_{\dot\beta} \tilde\phi^{\alpha(2+k)\dot\alpha(k)\dot\beta}
+ d_{2,k} e^{\alpha\dot\alpha} \phi^{\alpha(2+k)\dot\alpha(k-1)},
\nonumber \\
0 &=& D \tilde\phi^{\alpha(2+k)\dot\alpha(k+1)} + e_{\beta\dot\beta}
\tilde\phi^{\alpha(2+k)\beta\dot\alpha(k+1)\dot\beta} + c_{2,k}
e_\beta{}^{\dot\alpha} \phi^{\alpha(2+k)\beta\dot\alpha(k)} \\
 && + c_{1,k} e^\alpha{}_{\dot\beta} 
\bar\phi^{\alpha(1+k)\dot\alpha(1+k)\dot\beta} + d_{1,k}
e^{\alpha\dot\alpha} \tilde\phi^{\alpha(1+k)\dot\alpha(k)}, \nonumber
\end{eqnarray}
where
$$
c_{3,k} = - \frac{a_0}{2(k+3)(k+4)}, \qquad
c_{2,k} = - \frac{a_0}{(k+1)(k+2)}, \qquad
c_{1,k} = \frac{2\tilde{M}}{(k+2)(k+3)},
$$
$$
d_{2,k} = - \frac{\tilde{M}^2}{(k+1)(k+3)}, \qquad
d_{1,k} = - \frac{k(k+4)\tilde{M}^2}{(k+1)(k+2)^2(k+3)}.
$$

\subsection{Massive spin 2}

In this case we need both physical fields: one-forms
$H^{\alpha\dot\alpha}$, $A$ and zero-form
$\varphi$ as well as auxiliary fields\footnote{Recall that auxiliary
fields are fields that do not add any new physical degrees of freedom
(because their equations are algebraic) but necessary to bring the
free Lagrangian into first order form.}: one-forms 
$\Omega^{\alpha(2)}$, $\Omega^{\dot\alpha(2)}$ and zero-forms
$B^{\alpha(2)}$, $B^{\dot\alpha(2)}$ and $\pi^{\alpha\dot\alpha}$.
Gauge invariant curvatures for the physical fields look like:
\begin{eqnarray}
{\cal T}^{\alpha\dot\alpha} &=& D H^{\alpha\dot\alpha} + 
e_\beta{}^{\dot\alpha} \Omega^{\alpha\beta} + e^\alpha{}_{\dot\beta}
\Omega^{\dot\alpha\dot\beta} + M e^{\alpha\dot\alpha} A, \nonumber \\
{\cal A} &=& D A + 2(E_{\alpha(2)} B^{\alpha(2)} + E_{\dot\alpha(2)}
B^{\dot\alpha(2)}) + M e_{\alpha\dot\alpha} H^{\alpha\dot\alpha}, \\
\Phi &=& D \varphi + e_{\alpha\dot\alpha} \pi^{\alpha\dot\alpha}
+ M A, \nonumber
\end{eqnarray}
while for the auxiliary fields
\begin{eqnarray}
{\cal R}^{\alpha(2)} &=& D \Omega^{\alpha(2)} + M E^\alpha{}_\beta
B^{\alpha\beta} + \frac{M^2}{2} e^\alpha{}_{\dot\alpha}
H^{\alpha\dot\alpha} + 2M^2 E^{\alpha(2)} \varphi, \nonumber \\
{\cal B}^{\alpha(2)} &=& D B^{\alpha(2)} + M \Omega^{\alpha(2)} +
\frac{M}{2} e^\alpha{}_{\dot\alpha} \pi^{\alpha\dot\alpha}, \\
\Pi^{\alpha\dot\alpha}  &=& D \pi^{\alpha\dot\alpha} + M 
(e_\beta{}^{\dot\alpha} B^{\alpha\beta} + e^\alpha{}_{\dot\beta}
B^{\dot\alpha\dot\beta}) + M^2 H^{\alpha\dot\alpha} + M^2
e^{\alpha\dot\alpha} \varphi. \nonumber
\end{eqnarray}
Similarly to the well known "zero-torsion condition" in gravity, 
on-shell we put
$$
{\cal T}^{\alpha\dot\alpha} \approx 0, \qquad
{\cal A} \approx 0, \qquad \Phi \approx 0.
$$
As a result, the remaining curvatures satisfy a number of
differential
\begin{eqnarray}
D {\cal R}^{\alpha(2)} &\approx& M E^\alpha{}_\beta 
{\cal  B}^{\alpha\beta}, \nonumber \\
D {\cal B}^{\alpha(2)} &\approx& M {\cal R}^{\alpha(2)} - \frac{M}{2}
e^\alpha{}_{\dot\alpha} \Pi^{\alpha\dot\alpha}, \label{dif_2} \\
D \Pi^{\alpha\dot\alpha} &\approx& - M (e_\beta{}^{\dot\alpha} 
{\cal B}^{\alpha\beta} + e^\alpha{}_{\dot\beta} 
{\cal B}^{\dot\alpha\dot\beta}), \nonumber
\end{eqnarray}
as well as algebraic identities
\begin{equation}
e_\beta{}^{\dot\alpha} {\cal R}^{\alpha\beta} \approx 0, \qquad
E_{\alpha(2)} {\cal B}^{\alpha(2)} \approx 0, \qquad
e_{\alpha\dot\alpha} \Pi^{\alpha\dot\alpha} \approx 0. \label{alg_2}
\end{equation}
Unfolded equations for the auxiliary fields has the form
\begin{eqnarray}
0 &=& D \Omega^{\alpha(2)} + M E^\alpha{}_\beta B^{\alpha\beta}
+ \frac{M^2}{2} e^\alpha{}_{\dot\alpha} H^{\alpha\dot\alpha}
+ 2M^2 E^{\alpha(2)} \varphi + E_{\beta(2)} W^{\alpha(2)\beta(2)},
\nonumber \\
0 &=& D B^{\alpha(2)} + M \Omega^{\alpha(2)} + \frac{M}{2}
e^\alpha{}_{\dot\alpha} \pi^{\alpha\dot\alpha} + e_{\beta\dot\alpha}
B^{\alpha(2)\beta\dot\alpha}, \\
0 &=& D \pi^{\alpha\dot\alpha} + M (e_\beta{}^{\dot\alpha}
B^{\alpha\beta} + e^\alpha{}_{\dot\beta} B^{\dot\alpha\dot\beta})
+ M^2 H^{\alpha\dot\alpha} + M^2 e^{\alpha\dot\alpha} \varphi 
+ e_{\beta\dot\beta} \pi^{\alpha\beta\dot\alpha\dot\beta}. \nonumber
\end{eqnarray}
Here $W^{\alpha(4)}$, $B^{\alpha(3)\dot\alpha}$ and 
$\pi^{\alpha(2)\dot\alpha(2)}$ are the first representative of the
three infinite sets of gauge invariant zero-forms. Their equations
$(k \ge 0)$:
\begin{eqnarray}
0 &=& D W^{\alpha(4+k)\dot\alpha(k)} + e_{\beta\dot\beta}
W^{\alpha(4+k)\beta\dot\alpha(k)\dot\beta} + a_{3,k} 
e^\alpha{}_{\dot\beta} B^{\alpha(3+k)\dot\alpha(k)\dot\beta}
+ b_{2,k} e^{\alpha\dot\alpha} W^{\alpha(3+k)\dot\alpha(k-1)},
\nonumber \\
0 &=& D B^{\alpha(3+k)\dot\alpha(k+1)} +  e_{\beta\dot\beta}
B^{\alpha(3+k)\beta\dot\alpha(k+1)\dot\beta} + a_{2,k}
e_\beta{}^{\dot\alpha} W^{\alpha(3+k)\beta\dot\alpha(k)} \nonumber \\
 && + a_{1,k} e^\alpha{}_{\dot\beta} 
\pi^{\alpha(2+k)\dot\alpha(1+k)\dot\beta} + b_{1,k} 
e^{\alpha\dot\alpha} B^{\alpha(2+k)\dot\alpha(k)}, \\
0 &=& D \pi^{\alpha(2+k)\dot\alpha(2+k)} + e_{\beta\dot\beta}
\pi^{\alpha(2+k)\beta\dot\alpha(2+k)\dot\beta} + a_{0,k}
e_\beta{}^{\dot\alpha} B^{\alpha(2+k)\beta\dot\alpha(k+1)} \nonumber
 \\
 && + a_{0,k} e^\alpha{}_{\dot\beta} 
B^{\alpha(1+k)\dot\alpha(2+k)\dot\beta} + b_{0,k} e^{\alpha\dot\alpha}
\pi^{\alpha(1+k)\dot\alpha(1+k)}, \nonumber
\end{eqnarray}
where
$$
a_{3,k} = \frac{4M}{(k+4)(k+5)}, \qquad
a_{2,k} = \frac{M}{(k+1)(k+2)},
$$
$$ 
a_{1,k} = \frac{3M}{(k+3)(k+4)}, \qquad
a_{0,k} = \frac{2M}{(k+2)(k+3)}, 
$$
$$
b_{2,k} = - \frac{M^2}{(k+1)(k+4)}, \qquad
b_{0,k} = - \frac{k(k+5)M^2}{(k+2)^2(k+3)^2},
$$
$$
b_{1,k} = - \frac{k(k+5)M^2}{(k+1)(k+2)(k+3)(k+4)}.
$$

\subsection{Massive spin 5/2}
Here we need physical fields: one-forms $\Phi^{\alpha(2)\dot\alpha} +
h.c.$, $\Phi^\alpha + h.c.$ and zero-forms $\phi^\alpha + h.c.$ and
the so-called extra-fields\footnote{Recall that extra fields are the
fields that do not enter the free Lagrangian but necessary to
construct a complete set of the gauge invariant curvatures. The play 
an important role in the interacting theory.}: one-forms 
$\Phi^{\alpha(3)} + h.c.$ and zero-forms
$\phi^{\alpha(3)} + h.c.$, $\tilde\phi^{\alpha(2)\dot\alpha} + h.c.$.
Note that contrary to the spin 3/2 case zero-forms $\phi^{\alpha(3)}$
and $\tilde\phi^{\alpha(2)\dot\alpha}$ are now Stueckelberg fields and
not the gauge invariant ones. Gauge invariant curvatures for the
physical fields
\begin{eqnarray}
{\cal F}^{\alpha(2)\dot\alpha} &=& D \Phi^{\alpha(2)\dot\alpha} + 
e_\beta{}^{\dot\alpha} \Phi^{\alpha(2)\beta} + \frac{\tilde{M}}{2}
e^\alpha{}_{\dot\beta} \Phi^{\alpha\dot\alpha\dot\beta} 
+ \frac{a_0}{3} e^{\alpha\dot\alpha} \Phi^\alpha, \nonumber \\
{\cal F}^\alpha &=& D \Phi^\alpha + a_0 e_{\beta\dot\alpha}
\Phi^{\alpha\beta\dot\alpha} + \frac{3\tilde{M}}{2} 
e^\alpha{}_{\dot\alpha} \Phi^{\dot\alpha} 
- \frac{16\tilde{M}^2}{3} E^\alpha{}_\beta \phi^\beta
- 2a_0 E_{\beta(2)} \phi^{\alpha\beta(2)}, \\
{\cal C}^\alpha &=& D \phi^\alpha - \Phi^\alpha + \frac{3\tilde{M}}{2}
e^\alpha{}_{\dot\alpha} \phi^{\dot\alpha} + a_0 
e_{\beta\dot\alpha} \tilde\phi^{\alpha\beta\dot\alpha}, \nonumber
\end{eqnarray}
where
$$
a_0{}^2 = \frac{5\tilde{M}^2}{4},
$$
and for the extra fields
\begin{eqnarray}
{\cal F}^{\alpha(3)} &=& D \Phi^{\alpha(3)} + \frac{\tilde{M}^2}{3} 
e^\alpha{}_{\dot\alpha} \Phi^{\alpha(2)\dot\alpha} 
- \frac{2\tilde{M}^2}{3} E^\alpha{}_\beta \phi^{\alpha(2)\beta} 
- \frac{4\tilde{M}^2a_0}{9} E^{\alpha(2)} \phi^\alpha, \nonumber \\
{\cal C}^{\alpha(3)} &=& D \phi^{\alpha(3)} - \Phi^{\alpha(3)}
+ \frac{\tilde{M}^2}{3} e^\alpha{}_{\dot\alpha} 
\tilde\phi^{\alpha(2)\dot\alpha}, \\
{\cal C}^{\alpha(2)\dot\alpha} &=& D \phi^{\alpha(2)\dot\alpha} -
\Phi^{\alpha(2)\dot\alpha} + \frac{\tilde{M}}{2}  
e^\alpha{}_{\dot\beta} \bar\phi^{\alpha\dot\alpha\dot\beta} + 
e_\beta{}^{\dot\alpha} \phi^{\alpha(2)\beta} + \frac{a_0}{3}
e^{\alpha\dot\alpha} \phi^\alpha.
\nonumber
\end{eqnarray}
In this case on-shell means setting to zero curvatures for the
physical fields:
$$
{\cal F}^{\alpha(2)\dot\alpha} \approx 0, \qquad
{\cal F}^\alpha \approx 0, \qquad
{\cal C}^\alpha \approx 0.
$$
As a result, the remaining curvatures satisfy the following
differential
\begin{eqnarray}
D {\cal F}^{\alpha(3)} &\approx& - \frac{2\tilde{M}^2}{3} 
E^\alpha{}_\beta {\cal C}^{\alpha(2)\beta}, \nonumber \\
D {\cal C}^{\alpha(3)} &\approx& - {\cal F}^{\alpha(3)} - 
\frac{\tilde{M}^2}{3} e^\alpha{}_{\dot\alpha} 
{\cal C}^{\alpha(2)\dot\alpha}, \label{dif_3} \\
D {\cal C}^{\alpha(2)\dot\alpha} &\approx& - \frac{\tilde{M}}{2}
e^\alpha{}_{\dot\beta} {\cal C}^{\alpha\dot\alpha\dot\beta} -
e_\beta{}^{\dot\alpha} {\cal C}^{\alpha(2)\beta}, \nonumber
\end{eqnarray}
as well as algebraic identities
\begin{equation}
e_\beta{}^{\dot\alpha} {\cal F}^{\alpha(2)\beta} \approx 0, \qquad
E_{\beta(2)} {\cal C}^{\alpha\beta(2)} \approx 0, \qquad
e_{\beta\dot\alpha} {\cal C}^{\alpha\beta\dot\alpha} \approx 0.
\label{alg_3}
\end{equation}
Unfolded equations for the extra fields look like:
\begin{eqnarray}
0 &=& D \Phi^{\alpha(3)} + \frac{\tilde{M}^2}{3} 
e^\alpha{}_{\dot\alpha} \Phi^{\alpha(2)\dot\alpha} 
- \frac{2\tilde{M}^2}{3} E^\alpha{}_\beta \phi^{\alpha(2)\beta} 
- \frac{4\tilde{M}^2a_0}{9} E^{\alpha(2)} \phi^\alpha
 + E_{\beta(2)} Y^{\alpha(3)\beta(2)}, \nonumber \\
0 &=& D \phi^{\alpha(3)} - \Phi^{\alpha(3)}
+ \frac{\tilde{M}^2}{3} e^\alpha{}_{\dot\alpha} 
\tilde\phi^{\alpha(2)\dot\alpha} + e_{\beta\dot\alpha}
\phi^{\alpha(3)\beta\dot\alpha}, \\
0 &=& D \phi^{\alpha(2)\dot\alpha} - \Phi^{\alpha(2)\dot\alpha} +
\frac{\tilde{M}}{2}  e^\alpha{}_{\dot\beta}
\bar\phi^{\alpha\dot\alpha\dot\beta} +  e_\beta{}^{\dot\alpha}
\phi^{\alpha(2)\beta} + \frac{a_0}{3} e^{\alpha\dot\alpha} \phi^\alpha
+ e_{\beta\dot\beta} \tilde\phi^{\alpha(2)\beta\dot\alpha\dot\beta}.
\nonumber
\end{eqnarray}
Here $Y^{\alpha(5)}$, $\phi^{\alpha(4)\dot\alpha}$ and 
$\tilde\phi^{\alpha(3)\dot\alpha(2)}$ are the first representatives of
the three infinite sets of the gauge invariant zero-forms. Their
equations $(k \ge 0)$:
\begin{eqnarray}
0 &=& D Y^{\alpha(5+k)\dot\alpha(k)} + e_{\beta\dot\beta}
Y^{\alpha(5+k)\beta\dot\alpha(k)\dot\beta} + c_{5,k} 
e^\alpha{}_{\dot\beta} \phi^{\alpha(4+k)\dot\alpha(k)\dot\beta} 
+ d_{3,k} e^{\alpha\dot\alpha} Y^{\alpha(4+k)\dot\alpha(k-1)},
\nonumber \\
0 &=& D \phi^{\alpha(4+k)\dot\alpha(k+1)} + e_{\beta\dot\beta}
\phi^{\alpha(4+k)\beta\dot\alpha(k+1)\dot\beta} + c_{4,k}
e_\beta{}^{\dot\alpha} Y^{\alpha(4+k)\beta\dot\alpha(k)} \nonumber \\
 && + c_{3,k} e^\alpha{}_{\dot\beta} 
\tilde\phi^{\alpha(3+k)\dot\alpha(k+1)\dot\beta} + d_{2,k}
e^{\alpha\dot\alpha} \phi^{\alpha(3+k)\dot\alpha(k)}, \\
0 &=& D \tilde\phi^{\alpha(3+k)\dot\alpha(k+2)} + e_{\beta\dot\beta}
\tilde\phi^{\alpha(3+k)\beta\dot\alpha(k+2)\dot\beta} + c_{2,k}
e_\beta{}^{\dot\alpha} \phi^{\alpha(3+k)\beta\dot\alpha(k+1)}
\nonumber \\
 && + c_{1,k} e^\alpha{}_{\dot\beta} 
\bar{\phi}^{\alpha(2+k)\dot\alpha(k+2)\dot\beta} + d_{1,k}
e^{\alpha\dot\alpha} \tilde\phi^{\alpha(2+k)\dot\alpha(k+1)},
\nonumber
\end{eqnarray}
where
$$
c_{5,k} = - \frac{5M^2}{(k+5)(k+6)}, \qquad
c_{4,k} = - \frac{1}{(k+1)(k+2)}, \qquad
c_{3,k} = \frac{4M^2}{(k+4)(k+5)},
$$
$$
c_{2,k} = \frac{2}{(k+2)(k+3)}, \qquad
c_{1,k} = \frac{3M}{(k+3)(k+4)}, 
$$
$$
d_{3,k} = - \frac{M^2}{(k+1)(k+5)}, \qquad
d_{2,k} = - \frac{k(k+6)M^2}{(k+1)(k+2)(k+4)(k+5)},
$$
$$
d_{1,k} = - \frac{k(k+6)M^2}{(k+2)(k+3)^2(k+4)}.
$$

\subsection{Massive spin 3}

Here we have three sets of fields. Physical fields: one-forms
$H^{\alpha(2)\dot\alpha(2)}$, $H^{\alpha\dot\alpha}$, $A$ and
zero-form $\varphi$; auxiliary fields: one-forms 
$\Omega^{\alpha(3)\dot\alpha} + h.c.$, $\Omega^{\alpha(2)} +
h.c.$ and zero-forms $B^{\alpha(2)} + h.c.$, $\pi^{\alpha\dot\alpha}$;
extra fields: one-form $\Sigma^{\alpha(4)} + h.c.$ and zero-forms
$W^{\alpha(4)} + h.c.$, $B^{\alpha(3)\dot\alpha} + h.c.$,
$\pi^{\alpha(2)\dot\alpha(2)}$. Gauge invariant curvatures for the
physical fields
\begin{eqnarray}
{\cal T}^{\alpha(2)\dot{\alpha}(2)} &=& D 
H^{\alpha(2)\dot{\alpha}(2)} + e_\beta{}^{\dot{\alpha}}
\Omega^{\alpha(2)\beta\dot{\alpha}} + e^\alpha{}_{\dot{\beta}}
\Omega^{\alpha\dot{\alpha}(2)\dot{\beta}} + \frac{\rho_1}{2}
e^{\alpha\dot{\alpha}} H^{\alpha\dot{\alpha}}, \nonumber \\
{\cal T}^{\alpha\dot{\alpha}} &=& D H^{\alpha\dot{\alpha}} +
e_\beta{}^{\dot{\alpha}} \Omega^{\alpha\beta} + 
e^\alpha{}_{\dot{\beta}} \Omega^{\dot{\alpha}\dot{\beta}} + \rho_1
e_{\beta\dot{\beta}} H^{\alpha\beta\dot{\alpha}\dot{\beta}} +
\frac{\rho_0}{2} e^{\alpha\dot{\alpha}} A, \nonumber \\
{\cal A} &=& D A + \rho_0 e^{\alpha\dot{\alpha}} 
H_{\alpha\dot{\alpha}} - 2\rho_0 (E^{\alpha(2)} B_{\alpha(2)} +
E^{\dot{\alpha}(2)} B_{\dot{\alpha}(2)} ), \\
\Phi &=& D \varphi - A + \rho_0 e^{\alpha\dot{\alpha}}
\pi_{\alpha\dot{\alpha}}, \nonumber
\end{eqnarray}
where
$$
\rho_1{}^2 = \frac{M^2}{3}, \qquad \rho_0{}^2 = 5M^2,
$$
for the auxiliary fields
\begin{eqnarray}
{\cal R}^{\alpha(3)\dot{\alpha}} &=& D \Omega^{\alpha(3)\dot{\alpha}}
+ e_\beta{}^{\dot{\alpha}} \Sigma^{\alpha(3)\beta} + 
\frac{\rho_1}{4} e^{\alpha\dot{\alpha}} \Omega^{\alpha(2)} +
\frac{M^2}{6} e^\alpha{}_{\dot{\beta}} 
H^{\alpha(2)\dot{\alpha}\dot{\beta}}, \nonumber \\
{\cal R}^{\alpha(2)} &=& D \Omega^{\alpha(2)} + 3\rho_1
e_{\beta\dot{\alpha}} \Omega^{\alpha(2)\beta\dot{\alpha}} +
M^2 e^\alpha{}_{\dot{\alpha}} H^{\alpha\dot{\alpha}} \nonumber \\
 && - 6\rho_1 E_{\beta(2)} W^{\alpha(2)\beta(2)} - 
\frac{5M^2}{2} E^\alpha{}_\beta B^{\alpha\beta} - 
2M^2\rho_0 E^{\alpha(2)} \varphi, \\
{\cal B}^{\alpha(2)} &=& D B^{\alpha(2)} - \Omega^{\alpha(2)} +
3\rho_1 e_{\beta\dot{\alpha}} B^{\alpha(2)\beta\dot{\alpha}} +
M^2 e^\alpha{}_{\dot{\alpha}} \pi^{\alpha\dot{\alpha}}, \nonumber \\
\Pi^{\alpha\dot{\alpha}} &=& D \pi^{\alpha\dot{\alpha}} -
H^{\alpha\dot{\alpha}} + e_\beta{}^{\dot{\alpha}} B^{\alpha\beta} +
e^\alpha{}_{\dot{\beta}} B^{\dot{\alpha}\dot{\beta}} + \rho_1
e_{\beta\dot{\beta}} \pi^{\alpha\beta\dot{\alpha}\dot{\beta}} + 
\frac{\rho_0}{2} e^{\alpha\dot{\alpha}} \varphi \nonumber
\end{eqnarray}
and for the extra fields
\begin{eqnarray}
{\cal R}^{\alpha(4)} &=& D \Sigma^{\alpha(4)} + 
\frac{M^2}{4} e^\alpha{}_{\dot{\alpha}} 
\Omega^{\alpha(3)\dot{\alpha}} - \frac{M^2}{2} E^\alpha{}_\beta
W^{\alpha(3)\beta} - \frac{M^2\rho_1}{4} E^{\alpha(2)}
B^{\alpha(2)}, \nonumber \\
{\cal B}^{\alpha(4)} &=& D W^{\alpha(4)} - \Sigma^{\alpha(4)} +
\frac{M^2}{4} e^\alpha{}_{\dot{\alpha}} 
B^{\alpha(3)\dot{\alpha}}, \\
{\cal B}^{\alpha(3)\dot{\alpha}} &=& D B^{\alpha(3)\dot{\alpha}} -
\Omega^{\alpha(3)\dot{\alpha}} + e_\beta{}^{\dot{\alpha}}
W^{\alpha(3)\beta} + \frac{\rho_1}{4} e^{\alpha\dot{\alpha}}
B^{\alpha(2)} + \frac{M^2}{6} e^\alpha{}_{\dot{\beta}} 
\pi^{\alpha(2)\dot{\alpha}\dot{\beta}}, \nonumber \\
\Pi^{\alpha(2)\dot{\alpha}(2)} &=& D \pi^{\alpha(2)\dot{\alpha}(2)}
- H^{\alpha(2)\dot{\alpha}(2)} + e_\beta{}^{\dot{\alpha}}
B^{\alpha(2)\beta\dot{\alpha}} + e^\alpha{}_{\dot{\beta}}
B^{\alpha\dot{\alpha}(2)\dot{\beta}} + \frac{\rho_1}{2}
e^{\alpha\dot{\alpha}} \pi^{\alpha\dot{\alpha}}. \nonumber 
\end{eqnarray}
On-shell all curvatures for the physical and auxiliary fields are
zero, while the remaining ones satisfy:
\begin{eqnarray}
D {\cal R}^{\alpha(4)} &\approx& - \frac{M^2}{2} E^\alpha{}_\beta
{\cal B}^{\alpha(3)\beta}, \nonumber \\
D {\cal B}^{\alpha(4)} &\approx& - {\cal R}^{\alpha(4)} -
\frac{M^2}{4} e^\alpha{}_{\dot\alpha} {\cal B}^{\alpha(3)\dot\alpha},
\label{dif_4} \\
D {\cal B}^{\alpha(3)\dot\alpha} &\approx& - e_\beta{}^{\dot\alpha}
{\cal B}^{\alpha(3)\beta} - \frac{M^2}{6} e^\alpha{}_{\dot\beta}
\Pi^{\alpha(2)\dot\alpha\dot\beta}, \nonumber \\
D \Pi^{\alpha(2)\dot\alpha(2)} &\approx& - e_\beta{}^{\dot\alpha}
{\cal B}^{\alpha(2)\beta\dot\alpha} - e^\alpha{}_{\dot\beta}
{\cal B}^{\alpha\dot\alpha(2)\dot\beta}, \nonumber
\end{eqnarray}
\begin{equation}
e_\beta{}^{\dot\alpha} {\cal R}^{\alpha(3)\beta} \approx 0, \quad
E_{\beta(2)} {\cal B}^{\alpha(2)\beta(2)} \approx 0, \quad
e_{\beta\dot\alpha} {\cal B}^{\alpha(2)\beta\dot\alpha} \approx 0,
\quad e_{\beta\dot\beta} \Pi^{\alpha\beta\dot\alpha\dot\beta} \approx
0. \label{alg_4}
\end{equation}
Unfolded equations for the extra fields have the form
\begin{eqnarray}
0 &=& D \Sigma^{\alpha(4)} + \frac{M^2}{4} e^\alpha{}_{\dot\alpha} 
\Omega^{\alpha(3)\dot\alpha} - \frac{M^2}{2} E^\alpha{}_\beta
W^{\alpha(3)\beta} - \frac{M^2\rho_1}{4} E^{\alpha(2)} B^{\alpha(2)}
 + E_{\beta(2)} V^{\alpha(4)\beta(2)}, \nonumber \\
0 &=&  D W^{\alpha(4)} - \Sigma^{\alpha(4)} + \frac{M^2}{4} 
e^\alpha{}_{\dot\alpha} B^{\alpha(3)\dot\alpha}
 + e_{\beta\dot\alpha} W^{\alpha(4)\beta\dot\alpha}, \\
0 &=& D B^{\alpha(3)\dot\alpha} - \Omega^{\alpha(3)\dot\alpha} +
e_\beta{}^{\dot\alpha} W^{\alpha(3)\beta} + \frac{\rho_1}{4}
e^{\alpha\dot\alpha} B^{\alpha(2)} + \frac{M^2}{6} 
e^\alpha{}_{\dot\beta} \pi^{\alpha(2)\dot\alpha\dot\beta}
 + e_{\beta\dot\beta} B^{\alpha(3)\beta\dot\alpha\dot\beta}, \nonumber
\\
0 &=& D \pi^{\alpha(2)\dot\alpha(2)} - H^{\alpha(2)\dot\alpha(2)}
+ e_\beta{}^{\dot\alpha} B^{\alpha(2)\beta\dot\alpha} + 
e^\alpha{}_{\dot\beta} B^{\alpha\dot\alpha(2)\dot\beta} +
\frac{\rho_1}{2} e^{\alpha\dot\alpha} \pi^{\alpha\dot\alpha}
 + e_{\beta\dot\beta} \pi^{\alpha(2)\beta\dot\alpha(2)\dot\beta}.
\nonumber
\end{eqnarray}
Unfolded equations for the gauge invariant zero-forms
\begin{eqnarray}
0 &=& D V^{\alpha(6+k)\dot\alpha(k)} + e_{\beta\dot\beta}
V^{\alpha(6+k)\beta\dot\alpha(k)\dot\beta} + a_{5,k} 
e^\alpha{}_{\dot\beta} W^{\alpha(5+k)\dot\alpha(k)\dot\beta}
+ b_{3,k} e^{\alpha\dot\alpha} V^{\alpha(5+k)\dot\alpha(k-1)},
\nonumber \\
0 &=& D W^{\alpha(5+k)\dot\alpha(k+1)} + e_{\beta\dot\beta}
W^{\alpha(5+k)\beta\dot\alpha(k+1)\dot\beta} + a_{4,k} 
e_\beta{}^{\dot\alpha} V^{\alpha(5+k)\beta\dot\alpha(k)} \nonumber \\
 && + a_{3,k} e^\alpha{}_{\dot\beta}
B^{\alpha(4+k)\dot\alpha(k+1)\dot\beta} + b_{2,k} e^{\alpha\dot\alpha}
W^{\alpha(4+k)\dot\alpha(k)}, \nonumber \\
0 &=& D B^{\alpha(4+k)\dot\alpha(k+2)} + e_{\beta\dot\beta}
B^{\alpha(4+k)\beta\dot\alpha(k+2)\dot\beta} + a_{2,k}
e_\beta{}^{\dot\alpha} W^{\alpha(4+k)\beta\dot\alpha(k+1)} \\
 && + a_{1,k} e^\alpha{}_{\dot\beta} 
\pi^{\alpha(3+k)\dot\alpha(k+2)\dot\beta} + b_{1,k}
e^{\alpha\dot\alpha} B^{\alpha(3+k)\dot\alpha(k+1)}, \nonumber \\
0 &=& D \pi^{\alpha(3+k)\dot\alpha(k+3)} + e_{\beta\dot\beta}
\pi^{\alpha(3+k)\beta\dot\alpha(k+3)\dot\beta} + a_{0,k}
e_\beta{}^{\dot\alpha} B^{\alpha(3+k)\beta\dot\alpha(k+2)} \nonumber 
\\
 && + a_{0,k} e^\alpha{}_{\dot\beta} 
\bar{B}^{\alpha(2+k)\dot\alpha(k+3)\dot\beta} + b_{0,k}
e^{\alpha\dot\alpha} \pi^{\alpha(2+k)\dot\alpha(k+2)}, \nonumber 
\end{eqnarray}
where
$$
a_{5,k} = - \frac{6M^2}{(k+6)(k+7)}, \qquad
a_{4,k} = - \frac{1}{(k+1)(k+2)}, \qquad
a_{3,k} = \frac{5M^2}{(k+5)(k+6)},
$$
$$
a_{2,k} = \frac{2}{(k+2)(k+3)}, \qquad
a_{1.k} = \frac{2M^2}{(k+4)(k+5)}, \qquad
a_{0,k} = \frac{6}{(k+3)(k+4)},
$$
$$
b_{3,k} = - \frac{M^2}{(k+1)(k+6)}, \qquad
b_{2,k} = - \frac{k(k+7)M^2}{(k+1)(k+2)(k+5)(k+6)},
$$
$$
b_{1,k} = - \frac{k(k+7)M^2}{(k+2)(k+3)(k+4)(k+5)}, \qquad
b_{0,k} = - \frac{k(k+7)M^2}{(k+3)^2(k+4)^2}.
$$

\section{General analysis}

In this section we consider general properties for the cubic vertices
describing an interaction of the massless spin 3/2 gravitino and one
of the three massive superblocks $(2,3/2)$, $(5/2,2)$ or $(3,5/2)$.
Assuming that any such vertex can be reduced to the abelian form by
field redefinitions, we just consider the most general ansatz for
abelian vertices which do not vanish on-shell and require it to be
gauge invariant. In this, from the very beginning we will not assume
that bosonic and fermionic masses are equal. Remind, we work in four
dimensions and this severely restrict a number of abelian vertices 
that can be constructed in agreement with general results that in
$d=4$ there exist much less cubic vertices than in $d > 4$ (see e.g.
\cite{Vas11,Met22}). 

\subsection{Superblock $(2,3/2)$}

The most general ansatz for abelian vertices which do not vanish 
on-shell is
\begin{eqnarray}
{\cal L}_a &=& g_1 {\cal R}^{\alpha\beta} {\cal C}_\alpha \Psi_\beta
+ g_2 {\cal B}^{\alpha\beta} {\cal F}_\alpha \Psi_\beta + g_3
\Pi^{\alpha\dot\alpha} {\cal F}_{\dot\alpha} \Psi_\alpha \nonumber \\
 && + f_1 e_\alpha{}^{\dot\alpha} {\cal B}^{\alpha\beta}
{\cal C}_{\dot\alpha} \Psi_\beta + f_2 e^\alpha{}_{\dot\alpha}
{\cal B}^{\dot\alpha\dot\beta} {\cal C}_{\dot\beta} \Psi_\alpha
+ f_3 e^\alpha{}_{\dot\alpha} \Pi^{\beta\dot\alpha}
{\cal C}_\alpha \Psi_\beta + h.c. 
\end{eqnarray}
At the same time, we have two trivially gauge invariant vertices
\begin{equation}
{\cal L}_t = h_1 {\cal B}^{\alpha\beta} {\cal C}_\alpha 
D \Psi_\beta + h_2 \Pi^{\alpha\dot\alpha} 
{\cal C}_{\dot\alpha} D \Psi_\alpha + h.c. 
\end{equation}
Integrating by parts and using differential
(\ref{dif_1}), (\ref{dif_2}) and algebraic (\ref{alg_2}) on-shell
identities one can show that they are equivalent to particular
combinations of the abelian ones
\begin{eqnarray*}
h_1 \quad &\Rightarrow& \quad
g_1 = - M, \quad g_2 = - a_0, \quad g_3 = 0, \quad
f_1 = - \tilde{M}, \quad f_2 = 0, \quad f_3 = M, \\
h_2 \quad &\Rightarrow& \quad 
g_1 = g_2 = 0, \quad g_3 = - a_0, \quad
f_1 = f_2 = M, \quad f_3 = - \tilde{M}.
\end{eqnarray*}
Variations under the supertransformations produce
\begin{eqnarray}
\delta {\cal L}_a &=& [Mg_2 - a_0g_1] {\cal R}^{\alpha\beta}
{\cal F}_\alpha \zeta_\beta - [Mg_3 + a_0f_2] e^\alpha{}_{\dot\alpha}
{\cal B}^{\dot\alpha\dot\beta} {\cal F}_{\dot\beta} \zeta_\alpha
\nonumber \\
 && + [2Mg_1 + 2\tilde{M}f_1 + 2Mf_3 - \frac{a_0}{3}g_2]
E^\alpha{}_\gamma{\cal B}^{\beta\gamma} {\cal C}_\alpha \zeta_\beta
\nonumber \\
 && + [2Mf_1 + 2Mf_2 + 2\tilde{M}f_3 - \frac{a_0}{3}g_3]
E^\alpha{}_\beta \Pi^{\beta\dot\alpha} {\cal C}_{\dot\alpha}
\zeta_\alpha. 
\end{eqnarray}
For the both vertices $h_{1,2}$ these variations vanish and for
different masses they are the only solutions. Besides, there exists a
third solution
$$
Mf_1 = \tilde{M}f_3, \quad \tilde{M}f_1 = Mf_3
\quad \Rightarrow \quad M^2 = \tilde{M}^2
$$
with all other parameters being zero. In the unitary gauge it gives
\begin{equation}
{\cal L}_1 \sim [e_\alpha{}^{\dot\alpha} \Omega^{\alpha\beta}
\Phi_{\dot\alpha} + M e^\alpha{}_{\dot\alpha} H^{\beta\dot\alpha}
\Phi_\alpha ] \Psi_\beta + h.c. 
\end{equation}
Note that the vertex $h_2$ also has no more than one derivative in the
unitary gauge
\begin{equation}
\tilde{\cal L}_1 \sim H^{\alpha\dot\alpha} \Phi_{\dot\alpha} D
\Psi_\alpha + h.c. 
\end{equation}
Moreover, if we consider the minimal vertex (i.e. having no more than
one derivative) constructed in \cite{Zin18}, we will find that it
corresponds to some combination of these two vertices.

\subsection{Superblock $(5/2,2)$}

In this case an ansatz for the most general abelian vertices is
\begin{eqnarray}
{\cal L}_a &=& \Psi_\alpha [ g_1 {\cal R}_{\beta(2)}
{\cal C}^{\alpha\beta(2)} + g_2 {\cal R}_{\dot\alpha(2)}
{\cal C}^{\alpha\dot\alpha(2)} + g_3 {\cal B}_{\beta(2)}
{\cal F}^{\alpha\beta(2)} ] \nonumber \\
 && + \Psi_\alpha [ f_1 e^\alpha{}_{\dot\alpha} {\cal B}_{\beta(2)}
{\cal C}^{\beta(2)\dot\alpha} + f_2 e^\alpha{}_{\dot\alpha} 
{\cal B}_{\dot\beta(2)} {\cal C}^{\dot\alpha\dot\beta(2)} + f_3
e_\beta{}^{\dot\alpha} {\cal B}_{\dot\alpha\dot\beta} 
{\cal C}^{\alpha\beta\dot\beta}] \nonumber \\
 && + \Psi_\alpha [ f_4e_\beta{}^{\dot\alpha} \Pi_{\gamma\dot\alpha}
{\cal C}^{\alpha\beta\gamma} + f_5 e^\alpha{}_{\dot\alpha}
\Pi_{\beta\dot\beta} {\cal C}^{\beta\dot\alpha\dot\beta} ] + h.c.
\end{eqnarray}
while the trivially gauge invariant ones look like
\begin{equation}
{\cal L}_0 = D \Psi_\alpha [ h_1 {\cal B}_{\beta(2)} 
{\cal C}^{\alpha\beta(2)} + h_2 {\cal B}_{\dot\alpha(2)}
{\cal C}^{\alpha\dot\alpha(2)} + h_3 \Pi_{\beta\dot\alpha}
{\cal C}^{\alpha\beta\dot\alpha}] + h.c. 
\end{equation}
Here one also can show (with the help of identities (\ref{dif_2}),
(\ref{dif_3}) and (\ref{alg_2}), (\ref{alg_3})) that each one is
equivalent to some combination of the abelian vertices:
\begin{eqnarray*}
h_1 \quad &\Rightarrow& \quad
g_1 = M, \qquad g_3 = 1, \qquad 
f_1 = - \tilde{M}^2, \qquad f_4 = M, \\
h_2 \quad &\Rightarrow& \quad 
g_2 = M, \qquad f_2 = - 1, \qquad
f_3 = - \tilde{M}, \qquad f_5 = M, \\
h_3 \quad &\Rightarrow& \quad
f_1 = M, \qquad f_3 = M, \qquad
f_4 = - 1, \qquad f_5 = - \tilde{M}.
\end{eqnarray*}
Variations of the ansatz under the supertransformations produce
\begin{eqnarray}
\delta {\cal L}_a &=& [ g_1 - Mg_3 ] \eta_\alpha {\cal R}_{\beta(2)}
{\cal F}^{\alpha\beta(2)} + [g_2 + Mf_2   ]  \eta_\alpha 
e^\alpha{}_{\dot\alpha} {\cal R}_{\dot\beta(2)} 
{\cal C}^{\dot\alpha\dot\beta(2)} \nonumber \\
 && + [2Mg_1 - 2\tilde{M}^2g_3 - 2f_1 - 2Mf_4] \eta_\alpha 
E^\alpha{}_\beta {\cal B}_{\gamma(2)} {\cal C}^{\beta\gamma(2)}
\nonumber \\
 && + [ - 2Mg_2 - 2\tilde{M}^2f_2 + 2\tilde{M}f_3 + 2Mf_5] 
\eta_\alpha E^\alpha{}_\beta {\cal B}_{\dot\alpha(2)} 
{\cal C}^{\beta\dot\beta(2)} \nonumber \\
 && + [ 2Mf_1 - 2Mf_3 + 2\tilde{M}^2f_4 - 2\tilde{M}f_5] \eta_\alpha
E^\alpha{}_\beta \Pi_{\gamma\dot\alpha} 
{\cal C}^{\beta\gamma\dot\alpha}. 
\end{eqnarray}
For all three trivially invariant vertices these variations vanish.
There exists also a solution
$$
\tilde{M} f_3 + Mf_5 = 0, \qquad
Mf_3 + \tilde{M}f_5 = 0 \quad \Rightarrow \quad M^2 = \tilde{M}^2,
$$
with all other parameters being zero. In the unitary gauge it gives
\begin{equation}
{\cal L}_1 \sim \Psi_\alpha [ e_\beta{}^{\dot\alpha}
\Omega_{\dot\alpha\dot\beta} \Phi^{\alpha\beta\dot\beta} 
+ M e^\alpha{}_{\dot\alpha} H_{\beta\dot\beta} 
\Phi^{\beta\dot\alpha\dot\beta}] + h.c. 
\end{equation}
Note that among the trivially gauge invariant vertices only vertex
$h_3$ has no more that one derivative in the unitary gauge
\begin{equation}
\tilde{\cal L}_1 \sim D \Psi_\alpha H_{\beta\dot\alpha}
\Phi^{\alpha\beta\dot\alpha} + h.c. 
\end{equation}
By analogy with the previous case we may expect that the minimal
vertex corresponds to some combination of these two.

\subsection{Superblock $(3,5/2)$}

In this case the  most general ansatz for the abelian vertices is
\begin{eqnarray}
{\cal L}_1 &=& \Psi_\alpha [ g_1 {\cal R}^{\alpha\beta(3)}
{\cal C}_{\beta(3)} + g_2 {\cal B}^{\alpha\beta(3)}
{\cal F}_{\beta(3)} + g_3 {\cal B}^{\alpha\dot\alpha(3)}
{\cal F}_{\dot\alpha(3)} ] \nonumber \\
 && + \Psi_\alpha [ f_1 e_\beta{}^{\dot\alpha} 
{\cal B}^{\alpha\beta\gamma(2)} {\cal C}_{\gamma(2)\dot\alpha} 
+ f_2 e^\alpha{}_{\dot\alpha} {\cal B}^{\beta(3)\dot\alpha}
{\cal C}_{\beta(3)} + f_3 e_\beta{}^{\dot\alpha}
{\cal B}^{\alpha\beta\gamma\dot\beta} 
{\cal C}_{\gamma\dot\alpha\dot\beta} \nonumber \\
 && \qquad + f_4 e^\alpha{}_{\dot\alpha} 
\Pi^{\beta(2)\dot\alpha\dot\beta} {\cal C}_{\beta(2)\dot\beta} + f_5
e_\beta{}^{\dot\alpha} \Pi^{\alpha\beta\dot\beta(2)} 
{\cal C}_{\dot\alpha\dot\beta(2)} \nonumber  \\
 && \qquad + f_6 e^\alpha{}_{\dot\alpha} 
{\cal B}^{\beta\dot\alpha\dot\beta(2)} {\cal C}_{\beta\dot\beta(2)}  
 + f_7 e^\alpha{}_{\dot\alpha} {\cal B}^{\dot\alpha\dot\beta(3)}
{\cal C}_{\dot\beta(3)} ] + h.c.
\end{eqnarray}
while the trivially gauge invariant ones have the form
\begin{equation}
{\cal L}_0 = {\cal F}_\alpha [ h_1 {\cal B}^{\alpha\beta(3)}
{\cal C}_{\beta(3)} + h_2 {\cal B}^{\alpha\beta(2)\dot\alpha}
{\cal C}_{\beta(2)\dot\alpha} + h_3 \Pi^{\alpha\beta\dot\alpha(2)}
{\cal C}_{\beta\dot\alpha(2)} + h_4 {\cal B}^{\alpha\dot\alpha(3)}
{\cal C}_{\dot\alpha(3)}] + h.c.
\end{equation}
Here also one can show (with the help of identities (\ref{dif_3}),
(\ref{dif_4}) and (\ref{alg_3}), (\ref{alg_4})) that they are
equivalent to particular combinations of the abelian ones:
\begin{eqnarray*}
h_1 \quad &\Rightarrow& \quad
g_1 = - 1, \qquad g_2 = 1, \qquad 
f_1 = \tilde{M}^2, \qquad f_2 = - M^2, \\
h_2 \quad &\Rightarrow& \quad
f_1 = - 1, \qquad f_2 = 1, \qquad 
f_3 = \tilde{M}, \qquad f_4 = - \frac{M^2}{2}, \\
h_3 \quad &\Rightarrow& \quad
f_3 = - 2, \qquad f_4 = \tilde{M}, \qquad
f_5 = 1, \qquad f_6 = - 2, \\
h_4 \quad &\Rightarrow& \quad
g_3 = 1, \qquad f_5 = - \frac{M^2}{2}, \qquad 
f_6 = \tilde{M}^2, \qquad f_7 = - 1.
\end{eqnarray*}
Variations of the general ansatz under the supertransformations
produce
\begin{eqnarray}
\delta {\cal L}_a &=& [ g_1 + g_2] \eta_\alpha 
{\cal R}^{\alpha\beta(3)} {\cal F}_{\beta(3)}  + [g_3 + f_7 ]
e^\alpha{}_{\dot\alpha} {\cal B}^{\dot\alpha\dot\beta(3)} 
{\cal F}_{\dot\beta(3)} \nonumber \\
 && + [2M^2g_1 + 2\tilde{M}^2g_2 - 2f_1 - 2f_2 ] E^\alpha{}_\beta 
{\cal B}^{\beta\gamma(3)} {\cal C}_{\gamma(3)} \nonumber \\
 && + [ - 2\tilde{M}^2g_3 + 4f_5 + 2f_6 - 2M^2f_7] E^\alpha{}_\beta
{\cal B}^{\beta\dot\alpha(3)} {\cal C}_{\dot\alpha(3)} \nonumber \\
 && + [ 2M^2f_1 + 2\tilde{M}^2f_2 - 2\tilde{M}f_3 - 4f_4 ] 
E^\alpha{}_\beta {\cal B}^{\beta\gamma(2)\dot\alpha} 
{\cal C}_{\gamma(2)\dot\alpha} \nonumber \\
 && + [ M^2f_3 + 2\tilde{M}f_4 - 2\tilde{M}^2f_5 - M^2f_6 ]
E^\alpha{}_\beta \Pi^{\beta\gamma\dot\alpha(2)} 
{\cal C}_{\gamma\dot\alpha(2)}. 
\end{eqnarray}
For all four trivially invariant vertices these variations vanish.
There exists one more solution
$$
M^2 = \tilde{M}^2, \qquad f_6 = - 2f_5,
$$
with all other parameters being zero. In the unitary gauge it gives
\begin{equation}
{\cal L}_1 \sim \Psi_\alpha [ e^\alpha{}_{\dot\alpha}
\Omega^{\beta\dot\alpha\dot\beta(2)} \Phi_{\beta\dot\beta(2)}
+ M e_\beta{}^{\dot\alpha} H^{\alpha\beta\dot\beta(2)}
\Phi_{\dot\alpha\dot\beta(2)} ] + h.c. 
\end{equation}
Note that among the trivially gauge invariant vertices only the vertex
$h_3$ has no more than one derivative in the unitary gauge
\begin{equation}
\tilde{\cal L}_1 \sim D \Psi_\alpha H^{\alpha\beta\dot\alpha(2)}
\Phi_{\beta\dot\alpha(2)} + h.c. 
\end{equation}
Once again we have a combination of the minimal and non-minimal terms.
Moreover, it seems that these results can be generalized to the
arbitrary spins.

\section{Deformations for superblock $(2,3/2)$}

In this section we show (using massive superblock $(2,3/2)$ as an
example) how using unfolded equations one can fix the ambiguities
related with the field redefinitions. Let us take massive spin 3/2
and once again look at the equations for the gauge invariant
zero-forms:
\begin{eqnarray}
0 &=& D \phi^{\alpha(3+k)\dot\alpha(k)} + e_{\beta\dot\beta}
\phi^{\alpha(3+k)\beta\dot\alpha(k)\dot\beta} + c_{3,k}
e^\alpha{}_{\dot\beta} \tilde\phi^{\alpha(2+k)\dot\alpha(k)\dot\beta}
+ d_{2,k} e^{\alpha\dot\alpha} \phi^{\alpha(2+k)\dot\alpha(k-1)},
\nonumber \\
0 &=& D \tilde\phi^{\alpha(2+k)\dot\alpha(k+1)} + e_{\beta\dot\beta}
\tilde\phi^{\alpha(2+k)\beta\dot\alpha(k+1)\dot\beta} + c_{2,k}
e_\beta{}^{\dot\alpha} \phi^{\alpha(2+k)\beta\dot\alpha(k)} \\
 && + c_{1,k} e^\alpha{}_{\dot\beta} 
\bar\phi^{\alpha(1+k)\dot\alpha(1+k)\dot\beta} + d_{1,k}
e^{\alpha\dot\alpha} \tilde\phi^{\alpha(1+k)\dot\alpha(k)}. \nonumber
\end{eqnarray}
Algebraically, these equations allow one to express each zero-form as
the derivative of the previous one (plus low derivative corrections
proportional to $M$ and $M^2$). So these zero-forms describe all
higher derivatives of the physical field which do not vanish on-shell.
But these equations have also a geometrical meaning. Indeed, the
covariant derivative $D$ (containing background Lorentz connection)
and the background frame $e^{\alpha\dot\alpha}$ describe flat
Minkowsky space. They do it in a coordinate free way and all we need
for calculations are the two equations
\begin{equation}
D \wedge D = 0, \qquad D \wedge e^{\alpha\dot\alpha} = 0,
\end{equation}
where the first one means that curvature is zero while the second one
that torsion is zero. As a result these equations are invariant under
the global transformations corresponding to the Poincare algebra. The
explicit form of these transformations can be directly read out form
the equations and then one can calculate all commutators to check the
algebra. Now the idea is that to extend the Poincare algebra to super
Poincare we have to consider a deformation of these equations in the
presence of massless spin 3/2 gravitino. It must be possible simply
because flat Minkowsky space is not only solution of gravity equations
but also is a solution of supergravity equations. Naturally, for such
procedure to work we need some superpartner and we take massive spin 2
on this role. Note also that in this and following sections we assume
that bosonic and fermionic masses are equal.

\subsection{Spin 3/2}

Let us begin with the deformations for its gauge invariant zero-forms
\begin{eqnarray}
0 &=& D \phi^{\alpha(3+k)\dot\alpha(k)} + e_{\beta\dot\beta} 
\phi^{\alpha(3+k)\beta\dot\alpha(k)\dot\beta} + c_{3,k}
e^\alpha{}_{\dot\beta} \tilde\phi^{\alpha(2+k)\dot\alpha(k)\dot\beta}
+ d_{2,k} e^{\alpha\dot\alpha} \phi^{\alpha(2+k)\dot\alpha(k-1)}
\nonumber \\
 && + \gamma_3 W^{\alpha(3+k)\beta\dot\alpha(k)} \Psi_\beta
+ \delta_{3,k} W^{\alpha(3+k)\dot\alpha(k-1)} \Psi^{\dot\alpha}
\nonumber \\
 && + \gamma_2 B^{\alpha(3+k)\dot\alpha(k)\dot\beta} \Psi_{\dot\beta}
+ \delta_{2,k} B^{\alpha(2+k)\dot\alpha(k)} \Psi^\alpha, \nonumber \\
0 &=& D \tilde\phi^{\alpha(2+k)\dot\alpha(k+1)} + e_{\beta\dot\beta}
\tilde\phi^{\alpha(2+k)\beta\dot\alpha(k+1)\dot\beta} + c_{2,k} 
e_\beta{}^{\dot\alpha} \phi^{\alpha(2+k)\beta\dot\alpha(k)} \\
 && + c_{1,k} e^\alpha{}_{\dot\beta} 
\bar\phi^{\alpha(1+k)\dot\alpha(1+k)\dot\beta} + d_{1,k}
e^{\alpha\dot\alpha} \phi^{\alpha(1+k)\dot\alpha(k)} \nonumber \\
 && + \gamma_1 B^{\alpha(2+k)\beta\dot\alpha(k+1)} \Psi_\beta +
\delta_{1,k} B^{\alpha(2+k)\dot\alpha(k)} \Psi^{\dot\alpha} \nonumber
\\
 && + \gamma_0 \pi^{\alpha(2+k)\dot\alpha(k+1)\dot\beta}
\Psi_{\dot\beta} + \delta_{0,k} \pi^{\alpha(1+k)\dot\alpha(k+1)}
\Psi^\alpha.  \nonumber
\end{eqnarray}
Here zero-forms $W$, $B$ and $\pi$ come from the massive spin 2
equations. The consistency requirement leads to the unique solution:
$$
\gamma_2 = \gamma_3, \qquad
\gamma_1 = \gamma_0 = - \frac{a_0}{M}\gamma_3,
$$
$$
\delta_{3,k} = - \frac{M}{(k+1)}\gamma_3, \qquad
\delta_{2,k} = - \frac{kM}{(k+3)(k+4)}\gamma_3,
$$
$$
\delta_{1,k} = \frac{ka_0}{(k+1)(k+2)}\gamma_3, \qquad
\delta_{0,k} = \frac{ka_0}{(k+2)(k+3)}\gamma_3.
$$
Let us stress that the main reason for this is that these equations
form a closed subsystem and do not contain gauge and Stueckelberg
fields. From the other hand, the equations for gauge and Stueckelberg
fields do contain terms with the gauge invariant zero-forms and as a
result their deformations appear to be also unique. Indeed, let us
consider an ansatz for the Stueckelberg zero-form deformation
\begin{eqnarray}
0 &=& D \phi^\alpha - a_0 \Phi^\alpha + M e^\alpha{}_{\dot\alpha}
\phi^{\dot\alpha} + e_{\beta\dot\alpha}
\tilde\phi^{\alpha\beta\dot\alpha} \nonumber \\
 && + a_0a_1 B^{\alpha\beta} \Psi_\beta + a_0a_2
\pi^{\alpha\dot\alpha} \Psi_{\dot\alpha} + a_0a_3 \varphi \Psi^\alpha.
\end{eqnarray}
Consistency requires
$$
a_0a_1 = a_0a_2 = \gamma_1, \qquad a_3 = \frac{M}{2}a_1
$$
and in turn leads to
\begin{eqnarray}
0 &=& D \Phi^\alpha + M e^\alpha{}_{\dot\alpha}
\Phi^{\dot\alpha} - \frac{a_0}{3} E^\alpha{}_\beta \phi^\beta
+ E_{\beta(2)} \phi^{\alpha\beta(2)} \nonumber \\
 && + Ma_1 [\Omega^{\alpha\beta} \Psi_\beta + M
H^{\alpha\dot\alpha} \Psi_{\dot\alpha} + \frac{M}{2} A \Psi^\alpha
 +  e_\beta{}^{\dot\alpha} B^{\alpha\beta} \Psi_{\dot\alpha}
 + \frac{3M}{2} e^{\alpha\dot\alpha} \varphi \Psi_{\dot\alpha} ].
\end{eqnarray}
Now we return back to the Lagrangian formalism, i.e. we "forget" about
gauge invariant zero-forms and use the remaining terms as the
deformations of curvatures in the framework of Fradkin-Vasiliev
formalism:
\begin{eqnarray}
\Delta {\cal F}^\alpha &=& a_1 [ \Omega^{\alpha\beta} \Psi_\beta + M
H^{\alpha\dot\alpha} \Psi_{\dot\alpha} + \frac{M}{2} A \Psi^\alpha
+ e_\beta{}^{\dot\alpha} B^{\alpha\beta} \Psi_{\dot\alpha} +
\frac{3M}{2} e^{\alpha\dot\alpha} \varphi \Psi_{\dot\alpha}],
\nonumber \\
\Delta {\cal C}^\alpha &=& \frac{a_0}{M}a_1 [ B^{\alpha\beta}
\Psi_\beta + \pi^{\alpha\dot\alpha} \Psi_{\dot\alpha} + \frac{M}{2}
\varphi \Psi^\alpha ]. 
\end{eqnarray}
One can straightforwardly check that such deformations are consistent,
i.e. the deformed curvatures do transform covariantly. Note that 
these deformations determine not only supertransformations, but also
all other gauge transformations
\begin{equation}
\delta \Phi^\alpha = - a_1 [ \eta^{\alpha\beta} \Psi_\beta + M
\xi^{\alpha\dot\alpha} \Psi_{\dot\alpha} + \frac{M}{2} \xi
\Psi^\alpha],
\end{equation}
where $\eta^{\alpha(2)}$, $\xi^{\alpha\dot\alpha}$ and $\xi$ are the
gauge parameters of the massive spin 2.

\subsection{Spin 2}

Now we consider deformations for the massive spin 2 with the massive
spin 3/2 as a superpartner. The most general ansatz for its gauge
invariant zero-forms deformations has the form
\begin{eqnarray}
0 &=& D W^{\alpha(4+k)\dot\alpha(k)} + e_{\beta\dot\beta}
W^{\alpha(4+k)\beta\dot\alpha(k)\dot\beta} + a_{3,k} 
e^\alpha{}_{\dot\beta} B^{\alpha(3+k)\dot\alpha(k)\dot\beta}
+ b_{2,k} e^{\alpha\dot\alpha} W^{\alpha(3+k)\dot\alpha(k-1)}
\nonumber \\
 && + \alpha_3 \phi^{\alpha(4+k)\dot\alpha(k)\dot\beta} 
\Psi_{\dot\beta} + \beta_{3,k} \phi^{\alpha(3+k)\dot\alpha(k)}
\Psi^\alpha, \\
0 &=& D B^{\alpha(3+k)\dot\alpha(k+1)} +  e_{\beta\dot\beta}
B^{\alpha(3+k)\beta\dot\alpha(k+1)\dot\beta} + a_{2,k}
e_\beta{}^{\dot\alpha} W^{\alpha(3+k)\beta\dot\alpha(k)} \nonumber \\
 && + a_{1,k} e^\alpha{}_{\dot\beta} 
\pi^{\alpha(2+k)\dot\alpha(1+k)\dot\beta} + b_{1,k} 
e^{\alpha\dot\alpha} B^{\alpha(2+k)\dot\alpha(k)} \nonumber \\
 && + \alpha_2 \phi^{\alpha(3+k)\beta\dot\alpha(k+1)} \Psi_\beta
 + \beta_{2,k} \phi^{\alpha(3+k)\dot\alpha(k)} \Psi^{\dot\alpha}
\nonumber \\
 && + \alpha_1 \tilde\phi^{\alpha(3+k)\dot\alpha(k+1)\dot\beta}
\Psi_{\dot\beta} + \beta_{1,k} \tilde\phi^{\alpha(2+k)\dot\alpha(k+1)}
\Psi^\alpha,  \\
0 &=& D \pi^{\alpha(2+k)\dot\alpha(2+k)} + e_{\beta\dot\beta}
\pi^{\alpha(2+k)\beta\dot\alpha(2+k)\dot\beta} + a_{0,k}
e_\beta{}^{\dot\alpha} B^{\alpha(2+k)\beta\dot\alpha(k+1)} \nonumber 
\\
 && + a_{0,k} e^\alpha{}_{\dot\beta} 
B^{\alpha(1+k)\dot\alpha(2+k)\dot\beta} + b_{0,k} e^{\alpha\dot\alpha}
\pi^{\alpha(1+k)\dot\alpha(1+k)} \nonumber \\
 && + \alpha_0 \tilde\phi^{\alpha(2+k)\beta\dot\alpha(k+2)}
\Psi_\beta + \beta_{0,k} \tilde\phi^{\alpha(2+k)\dot\alpha(k+1)}
\Psi^{\dot\alpha} \nonumber \\
 && + \alpha_0 \bar\phi^{\alpha(k+2)\dot\alpha(k+2)\dot\beta}
\Psi_{\dot\beta} + \beta_{0,k} \bar\phi^{\alpha(k+1)(\dot\alpha(k+2)}
\Psi^\alpha. 
\end{eqnarray}
Here the consistency requirement also gives the unique solution:
$$
\alpha_2 = \frac{\alpha_3}{4}, \qquad
\alpha_1 = - \frac{a_0}{8M}\alpha_3, \qquad
\alpha_0 = - \frac{a_0}{12M}\alpha_3,
$$
$$
\beta_{3,k} = \frac{1}{3(k+4)}M\alpha_3, \qquad
\beta_{2,k} = \frac{(k+5)}{4(k+1)(k+2)}M\alpha_3,
$$
$$
\beta_{1,k} = - \frac{(k+5)}{8(k+3)(k+4)}a_0\alpha_3, \qquad
\beta_{0,k} = - \frac{(k+5)}{12(k+2)(k+3)}a_0\alpha_3.
$$
Now we proceed with the ansatz for the Stueckelberg zero-form
deformations:
\begin{eqnarray}
0 &=& D B^{\alpha(2)} + M \Omega^{\alpha(2)} + \frac{M}{2}
e^\alpha{}_{\dot\alpha} \pi^{\alpha\dot\alpha} +  e_{\beta\dot\alpha}
B^{\alpha(2)\beta\dot\alpha} \nonumber \\
 && + b_1 \phi^\alpha \Psi^\alpha + g_1 \phi^{\alpha(2)\beta}
\Psi_\beta + g_2 \tilde\phi^{\alpha(2)\dot\alpha} \Psi_{\dot\alpha}, 
\\
0 &=& D \pi^{\alpha\dot\alpha} + M (e_\beta{}^{\dot\alpha}
B^{\alpha\beta} + e^\alpha{}_{\dot\beta} B^{\dot\alpha\dot\beta})
+ M^2 H^{\alpha\dot\alpha} + M^2 e^{\alpha\dot\alpha} \varphi
+ e_{\beta\dot\beta} \pi^{\alpha\beta\dot\alpha\dot\beta} \nonumber \\
 && + b_2 (\phi^\alpha \Psi^{\dot\alpha} + \phi^{\dot\alpha}
\Psi^\alpha) + g_3 (\tilde\phi^{\alpha\beta\dot\alpha} \Psi_\beta 
+ \tilde\phi^{\alpha\dot\alpha\dot\beta} \Psi_{\dot\beta}), \\
0 &=& D \varphi + e_{\alpha\dot\alpha} \pi^{\alpha\dot\alpha}
+ M A \nonumber \\
 && + b_3 (\phi^\alpha \Psi_\alpha + \phi^{\dot\alpha}
\Psi_{\dot\alpha}). 
\end{eqnarray}
We obtain
$$
g_1 = - \frac{a_0}{2M}\alpha_0, \qquad
g_2 = \frac{3}{2}\alpha_0, \qquad
g_3 = \alpha_0,
$$
$$
b_1 = M\alpha_0, \qquad
b_2 = 2M\alpha_0, \qquad
b_3 = \alpha_0
$$
and this in turn leads to the following deformations for the gauge
fields:
\begin{eqnarray}
0 &=& D \Omega^{\alpha(2)} + a_0\alpha_0 \Phi^\alpha \Psi^\alpha
+ M\alpha_0 e^\alpha{}_{\dot\alpha} \phi^\alpha \Psi^{\dot\alpha}
- a_0\alpha_0 e_{\beta\dot\alpha} \phi^{\alpha(2)\beta}
\Psi^{\dot\alpha}, \nonumber \\
0 &=& D H^{\alpha\dot\alpha} + \frac{2a_0}{M}\alpha_0 (\Phi^\alpha
\Psi^{\dot\alpha} + \Phi^{\dot\alpha} \Psi^\alpha), \\
0 &=& D A + \frac{a_0}{M}\alpha_0 (\Phi^\alpha \Psi_\alpha +
\Phi^{\dot\alpha} \Psi_{\dot\alpha}) + 3\alpha_0 e_{\alpha\dot\alpha}
(\phi^\alpha \Psi^{\dot\alpha} + \phi^{\dot\alpha} \Psi^\alpha).
\nonumber
\end{eqnarray}
Thus we obtain the following deformations for all the gauge invariant
massive spin 2 curvatures:
\begin{eqnarray}
\Delta {\cal R}^{\alpha(2)} &=& \alpha_0 [M \Phi^\alpha \Psi^\alpha +
\frac{a_0}{6} e^\alpha{}_{\dot\alpha} \phi^\alpha \Psi^{\dot\alpha}],
 \nonumber \\
\Delta {\cal T}^{\alpha\dot\alpha} &=& 2\alpha_0 
(\Phi^\alpha \Psi^{\dot\alpha} + \Phi^{\dot\alpha} \Psi^\alpha ),
\nonumber \\
\Delta {\cal B}^{\alpha(2)} &=& \frac{a_0}{6}\alpha_0 \phi^\alpha
\Psi^\alpha,  \\
\Delta {\cal A} &=& \alpha_0 [ (\Phi^\alpha \Psi_\alpha +
\Phi^{\dot\alpha} \Psi_{\dot\alpha}) + \frac{a_0}{2M}
e_{\alpha\dot\alpha} (\phi^\alpha \Psi^{\dot\alpha} +
\phi^{\dot\alpha} \Psi^\alpha) ], \nonumber \\
\Delta \Pi^{\alpha\dot\alpha} &=& \frac{a_0}{3}\alpha_0 
(\phi^\alpha \Psi^{\dot\alpha} + \phi^{\dot\alpha} \Psi^\alpha),
\nonumber \\
\Delta \Phi &=& \frac{a_0}{6M}\alpha_0 (\phi^\alpha \Psi_\alpha
+ \phi^{\dot\alpha} \Psi_{\dot\alpha}). \nonumber 
\end{eqnarray}

\section{Deformations for superblock $(5/2,2)$}

In this section we apply the same procedure for the massive superblock
$(5/2,2)$.

\subsection{Spin 2}

Here an ansatz for the gauge invariant zero-forms deformations is
\begin{eqnarray}
0 &=& D W^{\alpha(4+k)\dot\alpha(k)} + e_{\beta\dot\beta}
W^{\alpha(4+k)\beta\dot\alpha(k)\dot\beta} + a_{3,k} 
e^\alpha{}_{\dot\beta} B^{\alpha(3+k)\dot\alpha(k)\dot\beta}
+ b_{2,k} e^{\alpha\dot\alpha} W^{\alpha(3+k)\dot\alpha(k-1)}
\nonumber \\
 && + \alpha_4 Y^{\alpha(4+k)\beta\dot\alpha(k)} \Psi_\beta
 + \beta_{4,k} Y^{\alpha(4+k)\dot\alpha(k-1)} \Psi^{\dot\alpha}
\nonumber \\
 && + \alpha_3 \phi^{\alpha(4+k)\dot\alpha(k)\dot\beta}
\Psi_{\dot\beta} + \beta_{3,k} \phi^{\alpha(3+k)\dot\alpha(k)}
\Psi^\alpha,  \\
0 &=& D B^{\alpha(3+k)\dot\alpha(k+1)} +  e_{\beta\dot\beta}
B^{\alpha(3+k)\beta\dot\alpha(k+1)\dot\beta} + a_{2,k}
e_\beta{}^{\dot\alpha} W^{\alpha(3+k)\beta\dot\alpha(k)} \nonumber \\
 && + a_{1,k} e^\alpha{}_{\dot\beta} 
\pi^{\alpha(2+k)\dot\alpha(1+k)\dot\beta} + b_{1,k} 
e^{\alpha\dot\alpha} B^{\alpha(2+k)\dot\alpha(k)} \nonumber \\
 && + \alpha_2 \phi^{\alpha(3+k)\beta\dot\alpha(k+1)} \Psi_\beta
 + \beta_{2,k} \phi^{\alpha(3+k)\dot\alpha(k)} \Psi^{\dot\alpha}
\nonumber \\
 && + \alpha_1 \tilde\phi^{\alpha(3+k)\dot\alpha(k+1)\dot\beta}
\Psi_{\dot\beta} + \beta_{1,k} 
\tilde\phi^{\alpha(2+k)\dot\alpha(k+1)} \Psi^\alpha,  \\
0 &=& D \pi^{\alpha(2+k)\dot\alpha(2+k)} + e_{\beta\dot\beta}
\pi^{\alpha(2+k)\beta\dot\alpha(2+k)\dot\beta} + a_{0,k}
e_\beta{}^{\dot\alpha} B^{\alpha(2+k)\beta\dot\alpha(k+1)} \nonumber
 \\
 && + a_{0,k} e^\alpha{}_{\dot\beta} 
B^{\alpha(1+k)\dot\alpha(2+k)\dot\beta} + b_{0,k} e^{\alpha\dot\alpha}
\pi^{\alpha(1+k)\dot\alpha(1+k)} \nonumber  \\
 && + \alpha_0 \tilde\phi^{\alpha(2+k)\beta\dot\alpha(k+2)} \Psi_\beta
+ \beta_{0,k} \tilde\phi^{\alpha(2+k)\dot\alpha(k+1)}
\Psi^{\dot\alpha} \nonumber \\
 && + \alpha_0 \bar\phi^{\alpha(k+2)\dot\alpha(k+2)\dot\beta}
\Psi_{\dot\beta} + \beta_{0,k} \bar\phi^{\alpha(k+1)\dot\alpha(k+2)}
\Psi^\alpha. 
\end{eqnarray}
Consistency requires
$$
\beta_{4,k} = \frac{1}{(k+1)}\beta_4, \qquad
\beta_{3,k} = \frac{k}{(k+4)(k+5)}\beta_3, \qquad
\beta_{2,k} = \frac{k}{(k+1)(k+2)}\beta_2, 
$$
$$
\beta_{1,k} = \frac{k}{(k+3)(k+4)}\beta_1, \qquad
\beta_{0,k} = \frac{k}{(k+2)(k+3)}\beta_0, 
$$
$$
\alpha_3 = \alpha_2 = - M\alpha_4, \qquad
\alpha_1 = \alpha_0 = - M^2\alpha_4.
$$
We proceed with the ansatz for Stueckelberg zero-forms deformations:
\begin{eqnarray}
0 &=& D B^{\alpha(2)} + m \Omega^{\alpha(2)} + \frac{M}{2}
e^\alpha{}_{\dot\alpha} \pi^{\alpha\dot\alpha} + e_{\beta\dot\alpha}
B^{\alpha(2)\beta\dot\alpha} \nonumber \\
 && + b_1 \phi^{\alpha(2)\beta} \Psi_\beta + b_2 
\tilde\phi^{\alpha(2)\dot\alpha} \Psi_{\dot\alpha} + b_3 \phi^\alpha
\Psi^\alpha, \\
0 &=& D \pi^{\alpha\dot\alpha} + M (e_\beta{}^{\dot\alpha}
B^{\alpha\beta} + e^\alpha{}_{\dot\beta} B^{\dot\alpha\dot\beta})
+ M^2 H^{\alpha\dot\alpha} + M^2 e^{\alpha\dot\alpha} \varphi 
+ e_{\beta\dot\beta} \pi^{\alpha\beta\dot\alpha\dot\beta} \nonumber \\
 && + b_4 (\tilde\phi^{\alpha\beta\dot\alpha} \Psi_\beta + h.c.)
 + b_5 (\phi^\alpha \Psi^{\dot\alpha} + h.c.), \\
0 &=& D \varphi + e_{\alpha\dot\alpha} \pi^{\alpha\dot\alpha}
+ M A \nonumber \\
 && + b_6 (\phi^\alpha \Psi_\alpha + h.c.).
\end{eqnarray}
Consistency gives
$$
b_1 = M\alpha_4, \qquad b_2 = b_4 = M^2\alpha_4, \qquad
b_3 = \frac{2Ma_0}{15}\alpha_4, \qquad
b_5 = \frac{2Ma_0}{5}\alpha_4, \qquad b_6 = \frac{4a_0}{5}\alpha_4
$$
and in turn leads to
\begin{eqnarray}
0 &=& D \Omega^{\alpha(2)} + M E^\alpha{}_\beta B^{\alpha\beta}
+ \frac{M^2}{2} e^\alpha{}_{\dot\alpha} H^{\alpha\dot\alpha}
+ 2M^2 E^{\alpha(2)} \varphi + E_{\beta(2)} W^{\alpha(2)\beta(2)}
\nonumber \\
 &&  + \alpha_4 [\Phi^{\alpha(2)\beta} \Psi_\beta + M
\Phi^{\alpha(2)\dot\alpha} \Psi_{\dot\alpha} +
\frac{2a_0}{15} \Phi^\alpha \Psi^\alpha
+ M e_{\beta\dot\alpha} \phi^{\alpha(2)\beta}
\Psi^{\dot\alpha} + \frac{8Ma_0}{15} e^\alpha{}_{\dot\alpha}
\phi^\alpha \Psi^{\dot\alpha}], \\
0 &=& D H^{\alpha\dot\alpha} + e_\beta{}^{\dot\alpha}
\Omega^{\alpha\beta} + e^\alpha{}_{\dot\beta}
\Omega^{\dot\alpha\dot\beta} + M e^{\alpha\dot\alpha} A \nonumber \\
 && + \alpha_4 [(\Phi^{\alpha\beta\dot\alpha} \Psi_\beta + h.c.) +
\frac{2a_0}{5M} (\Phi^\alpha \Psi^{\dot\alpha} + h.c.)],   \\
0 &=& D A + 2(E_{\alpha(2)} B^{\alpha(2)} + E_{\dot\alpha(2)}
B^{\dot\alpha(2)}) + M e_{\alpha\dot\alpha} H^{\alpha\dot\alpha}
\nonumber \\
 && + \alpha_4 [ \frac{M}{a_0} (\Phi^\alpha \Psi_\alpha + h.c.)
+ \frac{8a_0}{5}  e_{\alpha\dot\alpha} 
(\phi^\alpha \Psi^{\dot\alpha} + h.c. )] .
\end{eqnarray}
As a result we obtain the following consistent set of deformations for
all gauge invariant massive spin 2 curvatures
\begin{eqnarray}
\Delta {\cal R}^{\alpha(2)} &=& \alpha_4 [\Phi^{\alpha(2)\beta}
\Psi_\beta + M \Phi^{\alpha(2)\dot\alpha} \Psi_{\dot\alpha} +
\frac{2a_0}{15} \Phi^\alpha \Psi^\alpha
+ M e_{\beta\dot\alpha} \phi^{\alpha(2)\beta}
\Psi^{\dot\alpha} + \frac{8Ma_0}{15} e^\alpha{}_{\dot\alpha}
\phi^\alpha \Psi^{\dot\alpha}], \nonumber \\
\Delta {\cal T}^{\alpha\dot\alpha} &=& \alpha_4
[(\Phi^{\alpha\beta\dot\alpha} \Psi_\beta + h.c.) +
\frac{2a_0}{5M} (\Phi^\alpha \Psi^{\dot\alpha} + h.c.)], \nonumber \\
\Delta {\cal B}^{\alpha(2)} &=& \alpha_4 [ M \phi^{\alpha(2)\beta}
\Psi_\beta + M^2 \tilde\phi^{\alpha(2)\dot\alpha} \Psi_{\dot\alpha} +
\frac{2Ma_0}{15} \phi^\alpha \Psi^\alpha ], \\
\Delta {\cal A} &=& \alpha_4 [ \frac{M}{a_0} (\Phi^\alpha \Psi_\alpha
+ h.c.) + \frac{8a_0}{5}  e_{\alpha\dot\alpha} 
(\phi^\alpha \Psi^{\dot\alpha} + h.c. )], \nonumber \\
\Delta \Pi^{\alpha\dot\alpha} &=& \alpha_4 [ M^2
(\tilde\phi^{\alpha\beta\dot\alpha} \Psi_\beta + h.c.)
+ \frac{2Ma_0}{5} (\phi^\alpha \Psi^{\dot\alpha} + h.c.)], \nonumber 
\\
\Delta \Phi &=& \frac{4a_0}{5}\alpha_4 (\phi^\alpha \Psi_\alpha +
h.c.). \nonumber 
\end{eqnarray}

\subsection{Spin 5/2}

Similarly, we begin with the ansatz for gauge invariant zero-forms
deformations
\begin{eqnarray}
0 &=& D Y^{\alpha(5+k)\dot\alpha(k)} + e_{\beta\dot\beta}
Y^{\alpha(5+k)\beta\dot\alpha(k)\dot\beta} + c_{5,k} 
e^\alpha{}_{\dot\beta} \phi^{\alpha(4+k)\dot\alpha(k)\dot\beta} 
+ d_{3,k} e^{\alpha\dot\alpha} Y^{\alpha(4+k)\dot\alpha(k-1)}
\nonumber \\
 && + \gamma_4 W^{\alpha(5+k)\dot\alpha(k)\dot\beta} \Psi_{\dot\beta}
+ \delta_{4,k} W^{\alpha(4+k)\dot\alpha(k)} \Psi^\alpha, \\
0 &=& D \phi^{\alpha(4+k)\dot\alpha(k+1)} + e_{\beta\dot\beta}
\phi^{\alpha(4+k)\beta\dot\alpha(k+1)\dot\beta} + c_{4,k}
e_\beta{}^{\dot\alpha} Y^{\alpha(4+k)\beta\dot\alpha(k)} \nonumber \\
 && + c_{3,k} e^\alpha{}_{\dot\beta} 
\tilde\phi^{\alpha(3+k)\dot\alpha(k+1)\dot\beta} + d_{2,k}
e^{\alpha\dot\alpha} \phi^{\alpha(3+k)\dot\alpha(k)} \nonumber \\
 && + \gamma_3 W^{\alpha(4+k)\beta\dot\alpha(k+1)} \Psi_\beta +
\delta_{3,k} W^{\alpha(4+k)\dot\alpha(k)} \Psi^{\dot\alpha} \nonumber
\\
 && + \gamma_2 B^{\alpha(4+k)\dot\alpha(k+1)\dot\beta}
\Psi_{\dot\beta} + \delta_{2,k} B^{\alpha(3+k)\dot\alpha(k+1)}
\Psi^\alpha, \\
0 &=& D \tilde\phi^{\alpha(3+k)\dot\alpha(k+2)} + e_{\beta\dot\beta}
\tilde\phi^{\alpha(3+k)\beta\dot\alpha(k+2)\dot\beta} + c_{2,k}
e_\beta{}^{\dot\alpha} \phi^{\alpha(3+k)\beta\dot\alpha(k+1)}
\nonumber \\
 && + c_{1,k} e^\alpha{}_{\dot\beta} 
\bar{\phi}^{\alpha(2+k)\dot\alpha(k+2)\dot\beta} + d_{1,k}
e^{\alpha\dot\alpha} \tilde\phi^{\alpha(2+k)\dot\alpha(k+1)} \nonumber
\\
 && + \gamma_1 B^{\alpha(3+k)\beta\dot\alpha(k+2)} \Psi_\beta +
\delta_{1,k} B^{\alpha(3+k)\dot\alpha(k+1)} \Psi^{\dot\alpha}
\nonumber \\
 && + \gamma_0 \pi^{\alpha(3+k)\dot\alpha(k+2)\dot\beta} 
\Psi_{\dot\beta} + \delta_{0,k} \pi^{\alpha(2+k)\dot\alpha(k+2)}
\Psi^\alpha 
\end{eqnarray}
and obtain a solution
$$
\gamma_4 = - \frac{5M^2}{3}\gamma_0, \qquad
\gamma_3 = \frac{M}{3}\gamma_0, \qquad
\gamma_2 = \frac{4M}{3}\gamma_0, \qquad
\gamma_1 = \frac{2}{3}\gamma_0,
$$
$$
\delta_{2,k} = \frac{(k+6)}{(k+4)(k+5)}\delta_2, \qquad
\delta_{3,k} = \frac{(k+6)}{(k+1)(k+2)}\delta_3, \qquad
\delta_{4,k} = \frac{1}{(k+5)}\delta_4,
$$
$$
\delta_{0,k} = \frac{(k+6)}{(k+3)(k+4)}\delta_0, \qquad
\delta_{1,k} = \frac{(k+6)}{(k+2)(k+3)}\delta_1,
$$
$$
\delta_4 = - \frac{5M^2}{3}\delta_0, \qquad
\delta_3 = \frac{M}{3}\delta_0, \qquad
\delta_2 = \frac{4M}{3}\delta_0, \qquad
\delta_1 = \frac{2}{3}\delta_0, \qquad
\delta_0 = M\gamma_0.
$$
We proceed with the Stueckelberg zero-forms deformations
\begin{eqnarray}
0 &=& D \phi^{\alpha(3)} - \Phi^{\alpha(3)} + \frac{M^2}{3}
e^\alpha{}_{\dot\alpha} \tilde\phi^{\alpha(2)\dot\alpha} +
e_{\beta\dot\alpha} \phi^{\alpha(3)\beta\dot\alpha} \nonumber \\
 && + a_1 B^{\alpha(2)} \Psi^\alpha + g_1 W^{\alpha(3)\beta}
\Psi_\beta + g_2 B^{\alpha(3)\dot\alpha} \Psi_{\dot\alpha}, \\
0 &=& D \tilde\phi^{\alpha(2)\dot\alpha} - \Phi^{\alpha(2)\dot\alpha}
 + \frac{M}{2} e^\alpha{}_{\dot\beta}
\bar\phi^{\alpha\dot\alpha\dot\beta} + e_\beta{}^{\dot\alpha}
\phi^{\alpha(2)\beta} + \frac{a_0}{3} e^{\alpha\dot\alpha} \phi^\alpha
+ e_{\beta\dot\beta} \tilde\phi^{\alpha(2)\beta\dot\alpha\dot\beta}
\nonumber \\
 && + a_2 B^{\alpha(2)} \Psi^{\dot\alpha} + a_3 \pi^{\alpha\dot\alpha}
\Psi^\alpha + g_3 B^{\alpha(2)\beta\dot\alpha} \Psi_\beta + g_4
\pi^{\alpha(2)\dot\alpha\dot\beta} \Psi_{\dot\beta}, \\
0 &=& D \phi^\alpha - \Phi^\alpha + \frac{3M}{2}
e^\alpha{}_{\dot\alpha} \phi^{\dot\alpha} + a_0  e_{\beta\dot\alpha}
\tilde\phi^{\alpha\beta\dot\alpha} \nonumber \\
 && + a_4 B^{\alpha\beta} \Psi_\beta + a_5 \pi^{\alpha\dot\alpha}
\Psi_{\dot\alpha} + a_6 \varphi \Psi^\alpha.
\end{eqnarray}
We obtain
$$
g_1 = \frac{M}{3}\gamma_0, \qquad
g_2 = \frac{4M}{3}\gamma_0, \qquad
g_3 = \frac{2}{3}\gamma_0, \qquad
g_4 = \gamma_0,
$$
$$
a_1 = \frac{5M^2}{9}\gamma_0, \qquad
a_2 = \frac{5M}{3}\gamma_0, \qquad
a_3 = \frac{5M}{6}\gamma_0,
$$
$$
a_4 = \frac{2a_0}{3}\gamma_0, \qquad
a_5 = a_0\gamma_0, \qquad
a_6 = 2Ma_0 \gamma_0
$$
and for the gauge fields
\begin{eqnarray}
0 &=& D \Phi^{\alpha(3)}  + \frac{M^2}{3} e^\alpha{}_{\dot\alpha}
\Phi^{\alpha(2)\dot\alpha} - \frac{2M^2}{3} E^\alpha{}_\beta
\phi^{\alpha(2)\beta} - \frac{4M^2a_0}{9} E^{\alpha(2)} \phi^\alpha
 + E_{\beta(2)} Y^{\alpha(3)\beta(2)} \nonumber \\
 && + Ma_1 \Omega^{\alpha(2)} \Psi^\alpha - \frac{M^2}{3}a_2
B^{\alpha(2)} e^\alpha{}_{\dot\alpha} \Psi^{\dot\alpha}
 - \frac{M^2}{3}\gamma_0 e_{\beta\dot\alpha} W^{\alpha(3)\beta}
\Psi^{\dot\alpha} + \frac{M^2}{5}\gamma_0 e_{\beta\dot\alpha}
B^{\alpha(2)\beta\dot\alpha} \Psi^\alpha, \\
0 &=& D \Phi^{\alpha(2)\dot\alpha} + e_\beta{}^{\dot\alpha}
\Phi^{\alpha(2)\beta} + \frac{M}{2} e^\alpha{}_{\dot\beta}
\Phi^{\alpha\dot\alpha\dot\beta} + \frac{a_0}{3} e^{\alpha\dot\alpha}
\Phi^\alpha \nonumber \\
 && + Ma_2 \Omega^{\alpha(2)} \Psi^{\dot\alpha} + M^2a_3
H^{\alpha\dot\alpha} \Psi^\alpha,  \\
0 &=& D \Phi^\alpha + a_0 e_{\beta\dot\alpha}
\Phi^{\alpha\beta\dot\alpha} + \frac{3M}{2} e^\alpha{}_{\dot\alpha}
\Phi^{\dot\alpha} - \frac{16M^2}{3} E^\alpha{}_\beta \phi^\beta
- 2a_0 E_{\beta(2)} \phi^{\alpha\beta(2)} \nonumber \\
 && + Ma_4 \Omega^{\alpha\beta} \Psi_\beta + M^2a_5
H^{\alpha\dot\alpha} \Psi_{\dot\alpha} + Ma_6 A \Psi^\alpha
 - \frac{8Ma_0}{3}\gamma_0 B^{\alpha\beta} e_{\beta\dot\alpha}
\Psi^{\dot\alpha} + 4M^2\gamma_0 e^{\alpha\dot\alpha}
\varphi \Psi_{\dot\alpha}. 
\end{eqnarray}
Finally we get
\begin{eqnarray}
\Delta {\cal F}^{\alpha(3)} &=& a_2[\frac{M^2}{3} \Omega^{\alpha(2)}
\Psi^\alpha - \frac{M^2}{3} e^\alpha{}_{\dot\alpha} B^{\alpha(2)}
\Psi^{\dot\alpha}], \nonumber \\
\Delta {\cal C}^{\alpha(3)} &=& \frac{M}{3}a_2 B^{\alpha(2)}
\Psi^\alpha, \nonumber \\
\Delta {\cal F}^{\alpha(2)\dot\alpha(2)} &=& a_2[M \Omega^{\alpha(2)}
\Psi^{\dot\alpha} + \frac{M^2}{2} H^{\alpha\dot\alpha} \Psi^\alpha],
\nonumber \\
\Delta {\cal C}^{\alpha(2)\dot\alpha} &=& a_2 [B^{\alpha(2)}
\Psi^{\dot\alpha} + \frac{1}{2} \pi^{\alpha\dot\alpha} \Psi^\alpha], 
\\
\Delta {\cal F}^\alpha &=& a_2[\frac{2a_0}{5} \Omega^{\alpha\beta}
\Psi_\beta + \frac{3Ma_0}{5} H^{\alpha\dot\alpha} \Psi_{\dot\alpha}
+ \frac{6Ma_0}{5} A \Psi^\alpha \nonumber \\
 && \quad + \frac{8a_0}{5}  e_\beta{}^{\dot\alpha} B^{\alpha\beta}
\Psi_{\dot\alpha} + \frac{12M}{5} e^{\alpha\dot\alpha} \varphi
\Psi^\alpha], \nonumber \\
\Delta {\cal C}^\alpha &=& a_2[\frac{M}{2a_0} B^{\alpha\beta}
\Psi_\beta + \frac{3M}{4a_0} \pi^{\alpha\dot\alpha}
\Psi_{\dot\alpha} + \frac{6a_0}{5} \varphi \Psi^\alpha].  \nonumber 
\end{eqnarray}

\section{Deformations for superblock $(3,5/2)$}

Our last example --- massive superblock $(3,5/2)$.

\subsection{Spin 5/2}

Ansatz for the gauge invariant zero-forms deformations
\begin{eqnarray}
0 &=& D Y^{\alpha(5+k)\dot\alpha(k)} + e_{\beta\dot\beta}
Y^{\alpha(5+k)\beta\dot\alpha(k)\dot\beta} + c_{5,k} 
e^\alpha{}_{\dot\beta} \phi^{\alpha(4+k)\dot\alpha(k)\dot\beta} 
+ d_{3,k} e^{\alpha\dot\alpha} Y^{\alpha(4+k)\dot\alpha(k-1)}
\nonumber \\
 && + \gamma_5 V^{\alpha(5+k)\beta\dot\alpha(k)} \Psi_\beta
 + \delta_{5,k} V^{\alpha(5+k)\dot\alpha(k-1)} \Psi^{\dot\alpha}
\nonumber \\
 && + \gamma_4 W^{\alpha(5+k)\dot\alpha(k)\dot\beta} \Psi_{\dot\beta}
+ \delta_{4,k} W^{\alpha(4+k)\dot\alpha(k)} \Psi^\alpha, \\
0 &=& D \phi^{\alpha(4+k)\dot\alpha(k+1)} + e_{\beta\dot\beta}
\phi^{\alpha(4+k)\beta\dot\alpha(k+1)\dot\beta} + c_{4,k}
e_\beta{}^{\dot\alpha} Y^{\alpha(4+k)\beta\dot\alpha(k)} \nonumber \\
 && + c_{3,k} e^\alpha{}_{\dot\beta} 
\tilde\phi^{\alpha(3+k)\dot\alpha(k+1)\dot\beta} + d_{2,k}
e^{\alpha\dot\alpha} \phi^{\alpha(3+k)\dot\alpha(k)} \nonumber \\
 && + \gamma_3 W^{\alpha(4+k)\beta\dot\alpha(k+1)} \Psi_\beta
+ \delta_{3,k} W^{\alpha(4+k)\dot\alpha(k)} \Psi^{\dot\alpha}
\nonumber \\
 && + \gamma_2 B^{\alpha(4+k)\dot\alpha(k+1)\dot\beta}
\Psi_{\dot\beta} + \delta_{2,k} B^{\alpha(3+k)\dot\alpha(k+1)}
\Psi^\alpha, \\
0 &=& D \tilde\phi^{\alpha(3+k)\dot\alpha(k+2)} + e_{\beta\dot\beta}
\tilde\phi^{\alpha(3+k)\beta\dot\alpha(k+2)\dot\beta} + c_{2,k}
e_\beta{}^{\dot\alpha} \phi^{\alpha(3+k)\beta\dot\alpha(k+1)}
\nonumber \\
 && + c_{1,k} e^\alpha{}_{\dot\beta} 
\bar{\phi}^{\alpha(2+k)\dot\alpha(k+2)\dot\beta} + d_{1,k}
e^{\alpha\dot\alpha} \tilde\phi^{\alpha(2+k)\dot\alpha(k+1)} \nonumber
\\
 && + \gamma_1 B^{\alpha(3+k)\beta\dot\alpha(k+2)} \Psi_\beta
+ \delta_{1,k} B^{\alpha(3+k)\dot\alpha(k+1)} \Psi^{\dot\alpha}
\nonumber \\
 && + \gamma_0 \pi^{\alpha(3+k)\dot\alpha(k+2)\dot\beta}
\Psi_{\dot\beta} + \delta_{0,k} \pi^{\alpha(2+k)\dot\alpha(k+2)}
\Psi^\alpha. 
\end{eqnarray}
Consistency requires
$$
\gamma_4 = - M\gamma_5, \qquad
\gamma_3 = \gamma_1 = \gamma_5, \qquad
\gamma_2 = M\gamma_5, \qquad
\gamma_0 = \frac{M}{2}\gamma_5,
$$
$$
\delta_{5,k} = \frac{1}{(k+1)}\delta_5, \qquad
\delta_{4,k} = \frac{k}{(k+5)(k+6)}\delta_4, \qquad
\delta_{3,k} = \frac{k}{(k+1)(k+2)}\delta_3,
$$
$$
\delta_{2,k} = \frac{k}{(k+4)(k+5)}\delta_2, \qquad
\delta_{1,k} = \frac{k}{(k+2)(k+3)}\delta_1, \qquad
\delta_{0,k} = \frac{k}{(k+3)(k+4)}\delta_0.
$$
Then for the Stueckelberg zero-forms deformations
\begin{eqnarray}
0 &=& D \phi^{\alpha(3)} - \Phi^{\alpha(3)}
+ \frac{\tilde{M}^2}{3} e^\alpha{}_{\dot\alpha} 
\tilde\phi^{\alpha(2)\dot\alpha} + e_{\beta\dot\alpha}
\phi^{\alpha(3)\beta\dot\alpha} \nonumber \\
 && + a_1 W^{\alpha(3)\beta} \Psi_\beta + a_2 B^{\alpha(3)\dot\alpha}
\Psi_{\dot\alpha} + a_3 B^{\alpha(2)} \Psi^\alpha, \\
0 &=& D \phi^{\alpha(2)\dot\alpha} - \Phi^{\alpha(2)\dot\alpha} +
\frac{\tilde{M}}{2}  e^\alpha{}_{\dot\beta}
\bar\phi^{\alpha\dot\alpha\dot\beta} +  e_\beta{}^{\dot\alpha}
\phi^{\alpha(2)\beta} + \frac{a_0}{3} e^{\alpha\dot\alpha} \phi^\alpha
+ e_{\beta\dot\beta} \tilde\phi^{\alpha(2)\beta\dot\alpha\dot\beta}
\nonumber \\
 && + a_4 B^{\alpha(2)\beta\dot\alpha} \Psi_\beta  + a_5
\pi^{\alpha(2)\dot\alpha\dot\beta} \Psi_{\dot\beta} + a_6
B^{\alpha(2)} \Psi^{\dot\alpha} + a_7 \pi^{\alpha\dot\alpha}
\Psi^\alpha,  \\
0 &=& D \phi^\alpha - \Phi^\alpha + \frac{3\tilde{M}}{2}
e^\alpha{}_{\dot\alpha} \phi^{\dot\alpha} + a_0 
e_{\beta\dot\alpha} \tilde\phi^{\alpha\beta\dot\alpha} \nonumber \\
 && + a_8 B^{\alpha\beta} \Psi_\beta + a_9 \pi^{\alpha\dot\alpha}
\Psi_{\dot\alpha} + a_{10} \varphi \Psi^\alpha 
\end{eqnarray}
we obtain
$$
a_1 = \gamma_5, \qquad a_2 = Ma_1, \qquad
a_4 = a_1, \qquad a_5 = \frac{M}{2}a_1,
$$
$$
a_3 = \frac{\rho_1}{12}a_1, \qquad
a_6 = \frac{\rho_1}{2M}a_1, \qquad
a_7 = \frac{\rho_1}{4}a_1,
$$
$$
a_8 = \frac{a_0}{3\rho_1}a_1, \qquad
a_9 = \frac{Ma_0}{2\rho_1}a_1, \qquad
a_{10} = \frac{3\rho_1}{4}a_1.
$$
This in turn determines the deformations for the gauge field
equations:
\begin{eqnarray}
0 &=& D \Phi^{\alpha(3)} + \frac{M^2}{3} e^\alpha{}_{\dot\alpha}
\Phi^{\alpha(2)\dot\alpha} - \frac{2M^2}{3} E^\alpha{}_\beta
\phi^{\alpha(2)\beta} - \frac{4M^2a_0}{9} E^{\alpha(2)} \phi^\alpha
 + E_{\beta(2)} Y^{\alpha(3)\beta(2)}  \nonumber \\
 && - a_1 \Sigma^{\alpha(3)\beta} \Psi_\beta - a_2 \
\Omega^{\alpha(3)\dot\alpha} \Psi_{\dot\alpha} -  a_3
\Omega^{\alpha(2)} \Psi^\alpha - a_2 e_{\beta\dot\alpha}
W^{\alpha(3)\beta} \Psi^{\dot\alpha} - \frac{5M\rho_1}{12}a_1
e^\alpha{}_{\dot\alpha} B^{\alpha(2)} \Psi^{\dot\alpha}, \\
0 &=& D \Phi^{\alpha(2)\dot\alpha}  + e_\beta{}^{\dot\alpha}
\Phi^{\alpha(2)\beta} + \frac{M}{2} e^\alpha{}_{\dot\beta}
\Phi^{\alpha\dot\alpha\dot\beta} + \frac{a_0}{3} e^{\alpha\dot\alpha}
\Phi^\alpha  \nonumber \\
 &&  - a_4 \Omega^{\alpha(2)\beta\dot\alpha} \Psi_\beta - a_5
H^{\alpha(2)\dot\alpha\dot\beta} \Psi_{\dot\beta} - a_6
\Omega^{\alpha(2)} \Psi^{\dot\alpha} - a_7 H^{\alpha\dot\alpha}
\Psi^\alpha ,   \\
0 &=& D \Phi^\alpha + a_0 e_{\beta\dot\alpha}
\Phi^{\alpha\beta\dot\alpha} + \frac{3M}{2} e^\alpha{}_{\dot\alpha}
\Phi^{\dot\alpha} - \frac{16M^2}{3} E^\alpha{}_\beta \phi^\beta
- 2a_0 E_{\beta(2)} \phi^{\alpha\beta(2)} \nonumber \\
 && - a_8 \Omega^{\alpha\beta} \Psi_\beta - a_9 H^{\alpha\dot\alpha}
\Psi_{\dot\alpha} - a_{10} A \Psi^\alpha - \frac{2Ma_0}{3}a_1
e_{\beta\dot\alpha} B^{\alpha\beta} \Psi^{\dot\alpha}
- 3M\rho_1a_1 e^\alpha{}_{\dot\alpha}
\varphi \psi^{\dot\alpha}. 
\end{eqnarray} 
Thus we obtain the following set of consistent deformations for all
spin 5/2 curvatures:
\begin{eqnarray}
\Delta {\cal F}^{\alpha(3)} &=& - a_1 [\Sigma^{\alpha(3)\beta}
\Psi_\beta + M \Omega^{\alpha(3)\dot\alpha} \Psi_{\dot\alpha} + 
\frac{\rho_1}{2} \Omega^{\alpha(2)} \Psi^\alpha \nonumber \\
 && \qquad + M e_{\beta\dot\alpha} W^{\alpha(3)\beta}
\Psi^{\dot\alpha} + \frac{5M\rho_1}{12} e^\alpha{}_{\dot\alpha}
B^{\alpha(2)} \Psi^{\dot\alpha}], \nonumber \\
\Delta {\cal C}^{\alpha(3)} &=& a_1 [W^{\alpha(3)\beta} \Psi_\beta +
M B^{\alpha(3)\dot\alpha} \Psi_{\dot\alpha} + \frac{\rho_1}{12}
B^{\alpha(2)} \Psi^\alpha ], \nonumber \\
\Delta {\cal F}^{\alpha(2)\dot\alpha} &=& - a_1 [
\Omega^{\alpha(2)\beta\dot\alpha} \Psi_\beta + \frac{M}{2}
H^{\alpha(2)\dot\alpha\dot\beta} \Psi_{\dot\beta} +
\frac{\rho_1}{2m} \Omega^{\alpha(2)} \Psi^{\dot\alpha} +
\frac{\rho_1}{4} H^{\alpha\dot\alpha} \Psi^\alpha], \nonumber   \\
\Delta {\cal C}^{\alpha(2)\dot\alpha} &=& a_1 [
B^{\alpha(2)\beta\dot\alpha} \Psi_\beta  + \frac{M}{2}
\pi^{\alpha(2)\dot\alpha\dot\beta} \Psi_{\dot\beta} + 
\frac{\rho_1}{2m} B^{\alpha(2)} \Psi^{\dot\alpha} + 
\frac{\rho_1}{4} \pi^{\alpha\dot\alpha} \Psi^\alpha ], \\
\Delta {\cal F}^\alpha &=& - a_1 [ \frac{a_0}{3\rho_1}
\Omega^{\alpha\beta} \Psi_\beta + \frac{Ma_0}{2\rho_1}
H^{\alpha\dot\alpha} \Psi_{\dot\alpha} + \frac{3\rho_1}{4} A
\Psi^\alpha \nonumber \\
 && \qquad + \frac{2Ma_0}{3} e_{\beta\dot\alpha} B^{\alpha\beta}
\Psi^{\dot\alpha} + 3M\rho_1 e^\alpha{}_{\dot\alpha}
\varphi \Psi^{\dot\alpha}], \nonumber \\
\Delta {\cal C}^\alpha &=& a_1 [\frac{a_0}{3\rho_1} B^{\alpha\beta}
\Psi_\beta + \frac{Ma_0}{2\rho_1} \pi^{\alpha\dot\alpha}
\Psi_{\dot\alpha} + \frac{3\rho_1}{4} \varphi \Psi^\alpha ].
\nonumber 
\end{eqnarray}

\subsection{Spin 3}

Ansatz for gauge invariant zero-form deformations
\begin{eqnarray}
0 &=& D V^{\alpha(6+k)\dot\alpha(k)} + e_{\beta\dot\beta}
V^{\alpha(6+k)\beta\dot\alpha(k)\dot\beta} + a_{5,k} 
e^\alpha{}_{\dot\beta} W^{\alpha(5+k)\dot\alpha(k)\dot\beta}
+ b_{3,k} e^{\alpha\dot\alpha} V^{\alpha(5+k)\dot\alpha(k-1)}
\nonumber \\
 && + \alpha_5 Y^{\alpha(6+k)\dot\alpha(k)\dot\beta} \Psi_{\dot\beta}
+ \beta_{5,k} Y^{\alpha(5+k)\dot\alpha(k)} \Psi^\alpha, \\
0 &=& D W^{\alpha(5+k)\dot\alpha(k+1)} + e_{\beta\dot\beta}
W^{\alpha(5+k)\beta\dot\alpha(k+1)\dot\beta} + a_{4,k} 
e_\beta{}^{\dot\alpha} V^{\alpha(5+k)\beta\dot\alpha(k)} \nonumber \\
 && + a_{3,k} e^\alpha{}_{\dot\beta}
B^{\alpha(4+k)\dot\alpha(k+1)\dot\beta} + b_{2,k} e^{\alpha\dot\alpha}
W^{\alpha(4+k)\dot\alpha(k)} \nonumber \\
 && + \alpha_4 Y^{\alpha(5+k)\beta\dot\alpha(k+1)} \Psi_\beta
+ \beta_{4,k} Y^{\alpha(5+k)\dot\alpha(k)} \Psi^{\dot\alpha} \nonumber
\\
 && + \alpha_3 \phi^{\alpha(5+k)\dot\alpha(k+1)\dot\beta}
\Psi_{\dot\beta} + \beta_{3,k} \phi^{\alpha(4+k)\dot\alpha(k+1)}
\Psi^\alpha, \\
0 &=& D B^{\alpha(4+k)\dot\alpha(k+2)} + e_{\beta\dot\beta}
B^{\alpha(4+k)\beta\dot\alpha(k+2)\dot\beta} + a_{2,k}
e_\beta{}^{\dot\alpha} W^{\alpha(4+k)\beta\dot\alpha(k+1)} \nonumber 
\\
 && + a_{1,k} e^\alpha{}_{\dot\beta} 
\pi^{\alpha(3+k)\dot\alpha(k+2)\dot\beta} + b_{1,k}
e^{\alpha\dot\alpha} B^{\alpha(3+k)\dot\alpha(k+1)} \nonumber \\
 && + \alpha_2 \phi^{\alpha(4+k)\beta\dot\alpha(k+2)} \Psi_\beta
+ \beta_{2,k} \phi^{\alpha(4+k)\dot\alpha(k+1)} \Psi^{\dot\alpha}
\nonumber \\
 && + \alpha_1 \tilde\phi^{\alpha(4+k)\dot\alpha(k+2)\dot\beta}
\Psi_{\dot\beta} + \beta_{1,k} 
\tilde\phi^{\alpha(3+k)\dot\alpha(k+2)} \Psi^\alpha, \\
0 &=& D \pi^{\alpha(3+k)\dot\alpha(k+3)} + e_{\beta\dot\beta}
\pi^{\alpha(3+k)\beta\dot\alpha(k+3)\dot\beta} + a_{0,k}
e_\beta{}^{\dot\alpha} B^{\alpha(3+k)\beta\dot\alpha(k+2)} \nonumber 
\\
 && + a_{0,k} e^\alpha{}_{\dot\beta} 
\bar{B}^{\alpha(2+k)\dot\alpha(k+3)\dot\beta} + b_{0,k}
e^{\alpha\dot\alpha} \pi^{\alpha(2+k)\dot\alpha(k+2)} \nonumber \\
 && + \alpha_0 \tilde\phi^{\alpha(3+k)\beta\dot\alpha(k+3)} \Psi_\beta
+ \beta_{0,k} \tilde\phi^{\alpha(3+k)\dot\alpha(k+2)}
\Psi^{\dot\alpha} \nonumber \\
 && + \alpha_0 \bar\phi^{\alpha(3+k)\dot\alpha(k+3)\dot\beta}
\Psi_{\dot\beta} + \beta_{0,k} \bar\phi^{\alpha)2+k)\dot\alpha(k+3)}
\Psi^\alpha. 
\end{eqnarray}
Consistency requires
$$
\alpha_5 = - 6M\alpha_4, \quad
\alpha_3 = - 5M\alpha_4, \quad
\alpha_2 = - 2\alpha_4, \quad
\alpha_1 = - 4M\alpha_4, \quad
\alpha_0 = - 6\alpha_4,
$$
$$
\beta_{5,k} = \frac{1}{(k+6)}\beta_5, \qquad
\beta_{4,k} = \frac{(k+7)}{(k+1)(k+2)}\beta_4, \qquad
\beta_{3,k} = \frac{(k+7)}{(k+5)(k+6)}\beta_3,
$$
$$
\beta_{2,k} = \frac{(k+7)}{(k+2)(k+3)}, \qquad
\beta_{1,k} = \frac{(k+7)}{(k+4)(k+5)}\beta_1, \qquad
\beta_{0,k} = \frac{(k+7)}{(k+3)(k+4)}\beta_0, 
$$
$$
\beta_5 = - 6M^2\alpha_4, \quad
\beta_4 = M\alpha_4, \quad
\beta_3 = - 5M^2\alpha_4, \quad
\beta_2 = - 2M\alpha_4, \quad
\beta_1 = - 4M^2\alpha_4, \quad
\beta_0 = - 6M\alpha_4.
$$
Then for the Stueckelberg zero-form deformations
\begin{eqnarray}
0 &=&  D W^{\alpha(4)} - \Sigma^{\alpha(4)} + \frac{M^2}{4} 
e^\alpha{}_{\dot\alpha} B^{\alpha(3)\dot\alpha}
 + e_{\beta\dot\alpha} W^{\alpha(4)\beta\dot\alpha} \nonumber \\
 && + b_1 \phi^{\alpha(3)} \Psi^\alpha + g_1 Y^{\alpha(4)\beta}
\Psi_\beta + g_2 \phi^{\alpha(4)\dot\alpha} \Psi_{\dot\alpha}, \\
0 &=& D B^{\alpha(3)\dot\alpha} - \Omega^{\alpha(3)\dot\alpha} +
e_\beta{}^{\dot\alpha} W^{\alpha(3)\beta} + \frac{\rho_1}{4}
e^{\alpha\dot\alpha} B^{\alpha(2)} + \frac{M^2}{6} 
e^\alpha{}_{\dot\beta} \pi^{\alpha(2)\dot\alpha\dot\beta}
 + e_{\beta\dot\beta} B^{\alpha(3)\beta\dot\alpha\dot\beta} \nonumber
\\
 && + b_2 \phi^{\alpha(3)} \Psi^{\dot\alpha} + b_3
\tilde\phi^{\alpha(2)\dot\alpha} \Psi^\alpha + g_3 
\phi^{\alpha(3)\beta\dot\alpha} \Psi_\beta + g_4
\tilde\phi^{\alpha(3)\dot\alpha\dot\beta} \Psi_{\dot\beta},  \\
0 &=& D \pi^{\alpha(2)\dot\alpha(2)} - H^{\alpha(2)\dot\alpha(2)}
+ e_\beta{}^{\dot\alpha} B^{\alpha(2)\beta\dot\alpha} + 
e^\alpha{}_{\dot\beta} B^{\alpha\dot\alpha(2)\dot\beta} +
\frac{\rho_1}{2} e^{\alpha\dot\alpha} \pi^{\alpha\dot\alpha}
 + e_{\beta\dot\beta} \pi^{\alpha(2)\beta\dot\alpha(2)\dot\beta}
\nonumber \\
 && + b_4 \tilde\phi^{\alpha(2)\dot\alpha} \Psi^{\dot\alpha} 
+ g_5 \tilde\phi^{\alpha(2)\beta\dot\alpha(2)} \Psi_\beta,\\
0 &=& D B^{\alpha(2)} - \Omega^{\alpha(2)} +
3\rho_1 e_{\beta\dot{\alpha}} B^{\alpha(2)\beta\dot{\alpha}} +
M^2 e^\alpha{}_{\dot{\alpha}} \pi^{\alpha\dot{\alpha}} \nonumber \\
 && + b_5 \phi^{\alpha(2)\beta} \Psi_\beta + b_6
\tilde\phi^{\alpha(2)\dot\alpha} \Psi_{\dot\alpha} + b_7
\phi^\alpha \Psi^\alpha,  \\
0 &=& D \pi^{\alpha\dot{\alpha}} - H^{\alpha\dot{\alpha}} + 
e_\beta{}^{\dot{\alpha}} B^{\alpha\beta} +
e^\alpha{}_{\dot{\beta}} B^{\dot{\alpha}\dot{\beta}} + \rho_1
e_{\beta\dot{\beta}} \pi^{\alpha\beta\dot{\alpha}\dot{\beta}} + 
\frac{\rho_0}{2} e^{\alpha\dot{\alpha}} \varphi \nonumber \\
 && + b_8 (\tilde\phi^{\alpha\beta\dot\alpha} \Psi_\beta 
+ \bar\phi^{\alpha\dot\alpha\dot\beta} \Psi_{\dot\beta})
 + b_9 (\phi^\alpha \Psi^{\dot\alpha} + \phi^{\dot\alpha}
\Psi^\alpha),   \\
0 &=& D \varphi - A + \rho_0 e^{\alpha\dot\alpha}
\pi_{\alpha\dot\alpha} \nonumber \\
 && + b_{10} (\phi^\alpha \Psi_\alpha + \phi^{\dot\alpha}
\Psi_{\dot\alpha})
\end{eqnarray}
we obtain
$$
g_1 = \alpha_4, \qquad
g_2 = - 5M\alpha_4, \qquad
g_3 = - 2\alpha_4, \qquad 
g_4 = - 4M\alpha_4, \qquad
g_5 = - 6\alpha_4,
$$
$$
b_1 = \frac{M}{4\rho_1}b_5, \qquad
b_2 = b_4 = \frac{M}{\rho_1}{b_5}, \qquad
b_3 = \rho_1b_5, \qquad
b_5 = - 6\rho_1\alpha_4, \qquad
b_6 = 2Mb_5,
$$
$$
b_7 = \frac{4a_0}{3}b_5, \qquad
b_8 = b_5, \qquad 
b_9 = \frac{2a_0}{M}b_5, \qquad 
b_{10} = 2b_5. 
$$
At last, for the gauge field deformations we get
\begin{eqnarray}
0 &=& D \Sigma^{\alpha(4)} + \frac{M^2}{4} e^\alpha{}_{\dot\alpha} 
\Omega^{\alpha(3)\dot\alpha} - \frac{M^2}{2} E^\alpha{}_\beta
W^{\alpha(3)\beta} - \frac{M^2\rho_1}{4} E^{\alpha(2)} B^{\alpha(2)}
 + E_{\beta(2)} V^{\alpha(4)\beta(2)} \nonumber \\
 && - b_1 \Phi^{\alpha(3)} \Psi^\alpha - 6M\alpha_4
e_{\beta\dot\alpha} Y^{\alpha(4)\beta} \Psi^{\dot\alpha} 
- \frac{M^2}{4}b_2 e^\alpha{}_{\dot\alpha} \phi^{\alpha(3)}
\Psi^{\dot\alpha}, \\
0 &=& D \Omega^{\alpha(3)\dot\alpha} + e_\beta{}^{\dot{\alpha}}
\Sigma^{\alpha(3)\beta} + \frac{\rho_1}{4} e^{\alpha\dot{\alpha}}
\Omega^{\alpha(2)} + \frac{M^2}{6} e^\alpha{}_{\dot{\beta}} 
H^{\alpha(2)\dot{\alpha}\dot{\beta}} \nonumber \\
 && - b_2 \Phi^{\alpha(3)} \Psi^{\dot\alpha} - b_3
\Phi^{\alpha(2)\dot\alpha} \psi^\alpha,     \\
0 &=& D H^{\alpha(2)\dot\alpha(2)} + e_\beta{}^{\dot{\alpha}}
\Omega^{\alpha(2)\beta\dot{\alpha}} + e^\alpha{}_{\dot{\beta}}
\Omega^{\alpha\dot{\alpha}(2)\dot{\beta}} + \frac{\rho_1}{2}
e^{\alpha\dot{\alpha}} H^{\alpha\dot{\alpha}} \nonumber \\
 && - b_4 (\Phi^{\alpha(2)\dot\alpha} \Psi^{\dot\alpha} +
\Phi^{\alpha\dot\alpha(2)} \Psi^\alpha), \\
0 &=& D \Omega^{\alpha(2)} + 3\rho_1 e_{\beta\dot\alpha}
\Omega^{\alpha(2)\beta\dot\alpha} + M^2 e^\alpha{}_{\dot\alpha}
H^{\alpha\dot{\alpha}} \nonumber \\
 && - b_5 \Phi^{\alpha(2)\beta} \Psi_\beta - b_6 
\Phi^{\alpha(2)\dot\alpha} \Psi_{\dot\alpha} - b_7 \Phi^\alpha
\Psi^\alpha - 5Mb_5 e_{\beta\dot\alpha}
\phi^{\alpha(2)\beta} \Psi^{\dot\alpha} - \frac{8Ma_0}{3}b_5
e^\alpha{}_{\dot\alpha} \phi^\alpha \Psi^{\dot\alpha}, \\
0 &=& D H^{\alpha\dot\alpha} + e_\beta{}^{\dot{\alpha}}
\Omega^{\alpha\beta} + e^\alpha{}_{\dot{\beta}} 
\Omega^{\dot\alpha\dot\beta} + \rho_1 e_{\beta\dot{\beta}}
H^{\alpha\beta\dot{\alpha}\dot{\beta}} + \frac{\rho_0}{2}
e^{\alpha\dot{\alpha}} A \nonumber \\
 && - b_8 (\Phi^{\alpha\beta\dot\alpha} \Psi_\beta +
\Phi^{\alpha\dot\alpha\dot\beta} \Psi_{\dot\beta}) - b_9
(\Phi^\alpha \Psi^{\dot\alpha} + \Phi^{\dot\alpha} \Psi^\alpha), \\
0 &=& D A + \rho_0 e^{\alpha\dot{\alpha}} 
H_{\alpha\dot{\alpha}} - 2\rho_0 (E^{\alpha(2)} B_{\alpha(2)} +
E^{\dot{\alpha}(2)} B_{\dot{\alpha}(2)} ) \nonumber \\
 && - b_{10} (\Phi^\alpha \Psi_\alpha + \Phi^{\dot\alpha}
\Psi_{\dot\alpha}) - 8Mb_5 e_{\alpha\dot\alpha}
(\phi^\alpha \Psi^{\dot\alpha} + \phi^{\dot\alpha} \Psi^\alpha).
\end{eqnarray}
As a result we obtain a complete set of deformations for all gauge
invariant spin 3 curvatures:
\begin{eqnarray}
\Delta {\cal R}^{\alpha(4)} &=& - b_1 \Phi^{\alpha(3)} \Psi^\alpha
 - \frac{M^2}{4}b_2 e^\alpha{}_{\dot\alpha} \phi^{\alpha(3)}
\Psi^{\dot\alpha}, \nonumber \\
\Delta {\cal B}^{\alpha(4)} &=& b_1 \phi^{\alpha(3)} \Psi^\alpha,  
\nonumber \\
\Delta {\cal R}^{\alpha(3)\dot\alpha} &=& - b_2 \Phi^{\alpha(3)}
\Psi^{\dot\alpha} - b_3 \Phi^{\alpha(2)\dot\alpha} \Psi^\alpha,
\nonumber \\
\Delta {\cal B}^{\alpha(3)\dot\alpha} &=& b_1 \phi^{\alpha(3)}
\Psi^{\dot\alpha} + b_3 \tilde\phi^{\alpha(2)\dot\alpha} \Psi^\alpha,
\nonumber \\
\Delta {\cal T}^{\alpha(2)\dot\alpha(2)} &=& - b_4 
(\Phi^{\alpha(2)\dot\alpha} \Psi^{\dot\alpha} +
\Phi^{\alpha\dot\alpha(2)} \Psi^\alpha), \nonumber \\
\Delta \Pi^{\alpha(2)\dot\alpha(2)} &=& b_4 
(\tilde\phi^{\alpha(2)\dot\alpha} \Psi^{\dot\alpha} +
\tilde\phi^{\alpha\dot\alpha(2)} \Psi^\alpha),  \nonumber \\
\Delta {\cal R}^{\alpha(2)} &=& - b_5 \Phi^{\alpha(2)\beta} \Psi_\beta
- b_6 \Phi^{\alpha(2)\dot\alpha} \Psi_{\dot\alpha} - b_7 \Phi^\alpha
\Psi^\alpha \nonumber \\
 && \qquad - 5Mb_5 e_{\beta\dot\alpha}
\phi^{\alpha(2)\beta} \Psi^{\dot\alpha} - \frac{8Ma_0}{3}b_5
e^\alpha{}_{\dot\alpha} \phi^\alpha \Psi^{\dot\alpha}  \\
\Delta {\cal B}^{\alpha(2)} &=& b_5 \phi^{\alpha(2)\beta} \Psi_\beta
+ b_6 \tilde\phi^{\alpha(2)\dot\alpha} \Psi_{\dot\alpha} + b_7
\phi^\alpha \Psi^\alpha, \nonumber \\
\Delta {\cal T}^{\alpha\dot\alpha} &=& - b_8
(\Phi^{\alpha\beta\dot\alpha} \Psi_\beta +
\Phi^{\alpha\dot\alpha\dot\beta} \Psi_{\dot\beta}) - b_9
(\Phi^\alpha \Psi^{\dot\alpha} + \Phi^{\dot\alpha} \Psi^\alpha),
\nonumber \\
\Delta \Pi^{\alpha\dot\alpha} &=& b_8
(\tilde\phi^{\alpha\beta\dot\alpha} \Psi_\beta +
\tilde\phi^{\alpha\dot\alpha\dot\beta} \Psi_{\dot\beta})
+ b_9 (\phi^\alpha \Psi^{\dot\alpha} + \phi^{\dot\alpha}
\Psi^\alpha),  \nonumber \\
\Delta {\cal A} &=& - b_{10} (\Phi^\alpha \Psi_\alpha +
\Phi^{\dot\alpha} \Psi_{\dot\alpha}) - 8Mb_5
e_{\alpha\dot\alpha} (\phi^\alpha \Psi_{\dot\alpha} +
\phi^{\dot\alpha} \Psi^\alpha), \nonumber \\
\Delta \Phi &=& b_{10} (\phi^\alpha \Psi_\alpha + \phi^{\dot\alpha}
\Psi_{\dot\alpha}).  \nonumber
\end{eqnarray}

\section*{Conclusion}

The gauge invariant formulation for the massive higher spin fields
\cite{Zin01,Met06,Zin08b,PV10,KhZ19}  turned out to be useful for the
investigation of their different properties (including partially
massless and infinite spin limits) with the explicit construction of
massive. partially massless and infinite spin supermultiplets
\cite{Zin07a,BKhSZ19,BKhSZ19a,BKhSZ19b} being one of the non-trivial
examples. However, trying to apply this formalism for the construction
of the interactions one faces the ambiguity related with the possible
field redefinitions \cite{BDGT18}. The reason is the presence of the
so-called Stueckelberg fields with their non-homogeneous gauge
transformations. Happily, all these fields redefinitions do not change
the part of the Lagrangian which appears in the so-called unitary
gauge (when all Stueckelberg fields are set to zero) so we still have
some independent important results. Using the cubic interactions of
the massless spin 3/2 gravitino with the massive superblocks 
$(s_1,s_2)$, $s_1 - s_2 = 1/2$ as an example, we have shown how these
ambiguities can be used as one more way to classify possible
vertices. Note, that the same approach can be used for the cubic
vertices for the three massive fields as well. 

From the other hand, we have shown how using the so-called unfolded
formalism \cite{PV10,KhZ19,KhZ20} one can fix these ambiguities and
obtain consistent deformations for all gauge invariant curvatures
(both for the gauge fields as well as for the Stueckelberg ones)
which is the most important step in the Fradkin-Vasiliev formalism 
\cite{FV87,FV87a,Vas11}. Unfortunately, this works for the massive
fields only so that the way to construct deformations for the
massless field curvatures is still has to be found.

\end{document}